\documentclass[aps,superscriptaddress,floats,showpacs]{revtex4}
\usepackage{bm}
\usepackage{graphicx}
\usepackage{subfigure}

\begin{document}
\title{Aspect ratio dependence of heat transfer and large-scale flow in turbulent convection}
\author{Jorge Bailon-Cuba}
\affiliation{Institut f\"ur Thermo- und Fluiddynamik, Technische Universit\"at Ilmenau,
             Postfach 100565, D-98684 Ilmenau, Germany}
\author{Mohammad S. Emran\footnote{Corresponding author: mohammad.emran@tu-ilmen
au.de}}
\affiliation{Institut f\"ur Thermo- und Fluiddynamik, Technische Universit\"at Ilmenau,
             Postfach 100565, D-98684 Ilmenau, Germany}
\author{J\"org Schumacher}
\affiliation{Institut f\"ur Thermo- und Fluiddynamik, Technische Universit\"at Ilmenau,
             Postfach 100565, D-98684 Ilmenau, Germany}

\date{\today}

\begin{abstract}
The heat transport and corresponding changes in the large-scale circulation (LSC) in turbulent 
Rayleigh-B\'{e}nard convection are studied by means of three-dimensional direct 
numerical simulations as a function of the aspect ratio $\Gamma$ of a closed cylindrical 
cell and the Rayleigh number $Ra$.  The Prandtl number is $Pr=0.7$ throughout the study. 
The aspect ratio $\Gamma$  is varied between 0.5 and 12 for a Rayleigh number range 
between $10^7$ and $10^9$.  The Nusselt number $Nu$ is the dimensionless measure of the 
global turbulent heat transfer. For  small and moderate aspect ratios, the global heat transfer law 
$Nu=A\times Ra^{\beta}$ shows a power law dependence of both fit coefficients 
$A$ and $\beta$ on the aspect ratio.  A minimum of $Nu(\Gamma)$ is found at $\Gamma\approx 2.5$ 
and $\Gamma\approx 2.25$ for $Ra=10^7$ and $Ra=10^8$, respectively. This is the point where the LSC undergoes 
a transition from a single-roll to a double-roll pattern. With increasing aspect ratio, we detect complex 
multi-roll LSC configurations in the  convection cell. For larger aspect ratios $\Gamma\gtrsim 8$, our data  
indicate that the heat transfer becomes independent of the aspect ratio of the cylindrical cell. 
The aspect ratio dependence of the turbulent heat transfer for small and moderate $\Gamma$ is 
in line with a varying amount of energy contained in the LSC, as quantified by the Karhunen-Lo\`{e}ve or 
Proper Orthogonal Decomposition (POD) analysis 
of the turbulent convection field. The POD analysis  is conducted here by the snapshot method for at 
least 100 independent realizations of the turbulent fields. The primary POD mode, which replicates the 
time-averaged LSC patterns, transports about 50\% of the global heat for $\Gamma\ge 1$. The snapshot analysis 
enables a systematic disentanglement of the contributions of POD modes to the global turbulent heat transfer. 
Although the smallest scale -- the Kolmogorov scale $\eta_K$ -- and the largest scale -- the cell height 
$H$ -- are widely separated in a turbulent flow field, the LSC patterns in fully turbulent fields exhibit 
strikingly similar texture to those in the weakly nonlinear regime right above the onset of convection. 
Pentagonal or hexagonal circulation cells are observed preferentially if the aspect ratio is sufficiently large 
($\Gamma\gtrsim 8$).    
\end{abstract}

\maketitle

\section{Introduction}   
One of the most comprehensively studied turbulent flows is Rayleigh-B\'{e}nard 
convection, in which a complex three-dimensional turbulent motion is initiated by heating 
a fluid from below and cooling from above. Detailed measurements of the turbulent 
heat transport (e.g. Niemela {\it et al.} 2000, Funfschilling {\it et al.} 2005, Amati {\it et al.}
2005, Ahlers {\it et al.} 2009), the statistics of temperature fluctuations and their gradients 
(Castaing {\it et al.} 1989, Emran \& Schumacher 2008), and more recently, of coherent thermal 
plume structures (Zhou {\it et al.} 2007, Shishkina \& Wagner 2008), which carry the heat 
locally through the closed vessel, have been conducted. The variation of turbulent heat transfer 
with respect to two of the three dimensionless control parameters in thermal convection -- 
the Rayleigh number $Ra$ and the Prandtl number $Pr$ -- was the focus of most of the 
laboratory experiments and simulations. The dependence on the third control parameter,  
the aspect ratio $\Gamma=D/H$ with $D$ being the sidelength or diameter and $H$ the cell 
height, has been studied much less intensively. 

Only a few systematic analyses of high-Rayleigh-number convection in flat cells with $\Gamma>1$ 
have been reported (Fitzjarrald 1976, Wu \& Libchaber 1992, Funfschilling {\it et al.} 2005, Hartlep {\it et al.} 2005, 
Sun {\it et al.}  2005, Niemela \& Sreenivasan 2006, du Puits {\it et al.} 2007) although the 
large-aspect ratio setting is relevant for nearly all geophysical and astrophysical flows (e.g. 
Stein \& Nordlund 2006) and many technological applications such as the energy-efficient 
design of indoor ventilation (e.g. Zerihun Desta {\it et al.} 2005). Furthermore, an explicit 
dependence on the aspect ratio is not contained in any of the existing scaling theories for the 
turbulent heat transfer (Siggia 1994, Grossmann \& Lohse 2000). Grossmann \& Lohse 
(2003) discussed geometry effects by including variations of the kinetic boundary layer thickness 
at the plates and side walls as a function of the  aspect ratio. However, they found that the global heat transfer laws remained 
independent of $\Gamma$.  This is because their argumentation is built on the volume flux 
conservation which requires a large scale flow -- a  so-called ``wind of turbulence".  In fact all 
existing scaling theories require such a large-scale flow for the ansatz of their boundary layer 
dynamics. It is, however, well-known that the coherent large-scale circulations present at 
$\Gamma\sim 1$ break down into more complex and less coherent patterns when the aspect 
ratio is increased far beyond unity. Such phenomena were  reported by several authors: for example 
by means of Fourier spectrum analysis (Hartlep {\it et al.} 2003), plume structure visualizations 
(Shishkina \& Wagner 2006) or comparisons of the autocorrelations of the temperature and velocity 
fields (du Puits {\it et al.} 2007).   

In this work, we therefore want to study the dependence of convective turbulence on the aspect ratio
in a cylindrical cell by three-dimensional direct numerical simulations. Our focus is on aspect ratios 
$\Gamma$ larger than unity.  Values for $\Gamma$ cover a range  between 0.5 and 12 for 
Rayleigh numbers between $10^7$ and $10^9$. The Prandtl number is kept constant. The present 
analysis addresses the following three questions: Does the turbulent heat transfer at fixed Rayleigh 
and Prandtl numbers depend on the aspect ratio? Similar studies have been carried out by
Fitzjarald (1976), Wu \& Libchaber 
(1992), Funfschilling {\it et al.} (2005), Hartlep {\it et al.} (2005) and Sun {\it et al.}  (2005).  Which 
changes in the global flow structure are associated with an increase of the aspect ratio beyond unity?  This aspect has been discussed in part in Hartlep {\it et al.} (2003) and du Puits {\it et al.} (2007).  
Which fraction of the total kinetic energy is carried and how much heat is transferred by the large-scale 
circulation (LSC)? Answering the last question requires a systematic disentanglement of the 
turbulent large- and small-scale flow and temperature patterns. Therefore, we conduct Proper Orthogonal Decomposition (POD) of the turbulent flow fields. We observe a dependence of 
the heat transfer -- as measured by the dimensionless Nusselt number $Nu$-- on $\Gamma$. 
This dependence is due to the rearrangement of the large-scale flow patterns with varying aspect ratio. 

Although the flow is fully turbulent, the time-averaged velocity field patterns will exhibit morphological 
similarities with the structures which have been observed at the onset of convection 
(Charlson \& Sani 1971, Oresta {\it et al.} 2007) or in the weakly nonlinear regime right above the
onset of convection (see e.g. Busse \& Whitehead 1971, Croquette 1989, Clever \& Busse 1989, 
Bodenschatz {\it et al.} 2000). For example, a transition from a one-roll to a two-roll flow 
pattern right above the critical Rayleigh number was found at an aspect ratio $\Gamma=1.62$ in a cylindrical 
cell (Oresta {\it et al.} 2007). In a fully turbulent regime, such bifurcations are present at slightly larger aspect ratios.  
Furthermore, in the weakly nonlinear case, many specific configurations have been detected such as knot 
convection,  spiral defect chaos, and textures with wall foci (see Bodenschatz {\it et al.} (2000) for a review). 
We observe that the time-averaged flow fields for $Ra\ge 10^7$ yield similar patterns. Our findings will be in line with  
recent studies by Hartlep {\it et al.} (2005) in a  rectangular cell with periodic 
side walls, in which emphasis was given to the variation of patterns with respect to the Prandtl 
number $Pr$ at different aspect ratios. Large-scale flow patterns are also present in other closed turbulent flow systems,
such as in high-Reynolds number turbulence in von K\'{a}rman swirling flows (La Porta {\it et al.} 2001) or 
Taylor vortex flows (Lathrop {\it et al.} 1992).

The outline of the paper is as follows. In the next section, we summarise the numerical model
and the equations of motion. The subsequent section discusses the dependence of the Nusselt number
on the aspect ratio for fixed Rayleigh numbers and the Nusselt number as a function of the 
Rayleigh number at fixed aspect ratio. Section 4 studies the LSC. Section 5 describes the POD 
analysis and the contributions of different POD modes to the heat transfer. We conclude with a summary
and outlook.  

\section{Numerical model}
The Navier--Stokes equations for an incompressible flow in the 
Boussinesq approximation are solved in combination with the advection--diffusion equation
for the temperature field in cylindrical coordinates. The system is given by
\begin{eqnarray}
\label{nseq}
\frac{\partial{\bm u}}{\partial t}+({\bm u}\cdot{\bm \nabla}){\bm u}
&=&-{\bm \nabla} p+\nu {\bm \nabla}^2{\bm u}+\alpha g T {\bm e}_z\,,\\
\label{ceq}
{\bm \nabla}\cdot{\bm u}&=&0\,,\\
\frac{\partial T}{\partial t}+({\bm u}\cdot{\bm \nabla}) T
&=&\kappa {\bm \nabla}^2 T\,,
\label{pseq}
\end{eqnarray}
where $p({\bm x},t)$ is the pressure, ${\bm u}({\bm x},t)$ the velocity field, 
$T({\bm x},t)$ the total temperature field, $\nu$ the kinematic viscosity, and $\kappa$
the diffusivity of the temperature. Our studies are conducted for $Pr=\nu/\kappa=0.7$.
The Rayleigh numbers $Ra=\alpha g \Delta T H^3/(\nu \kappa)$ span a range from 
$10^7$ to $10^9$. Here, $\alpha$ is the thermal expansion coefficient, 
$g$ the gravitational acceleration, and $\Delta T$ the outer temperature difference. The horizontal 
plates have no-slip boundary conditions, i.e., ${\bm u}\equiv 0$, at a fixed temperature. The side 
walls are adiabatic no-slip boundaries, i.e.,  ${\bm u}\equiv 0$ and $\partial T/\partial r=0$. Small 
effects of finite conductivity -- as present in the experiments -- are thus excluded. 
\begin{figure}
\begin{center}
\includegraphics[width=11cm]{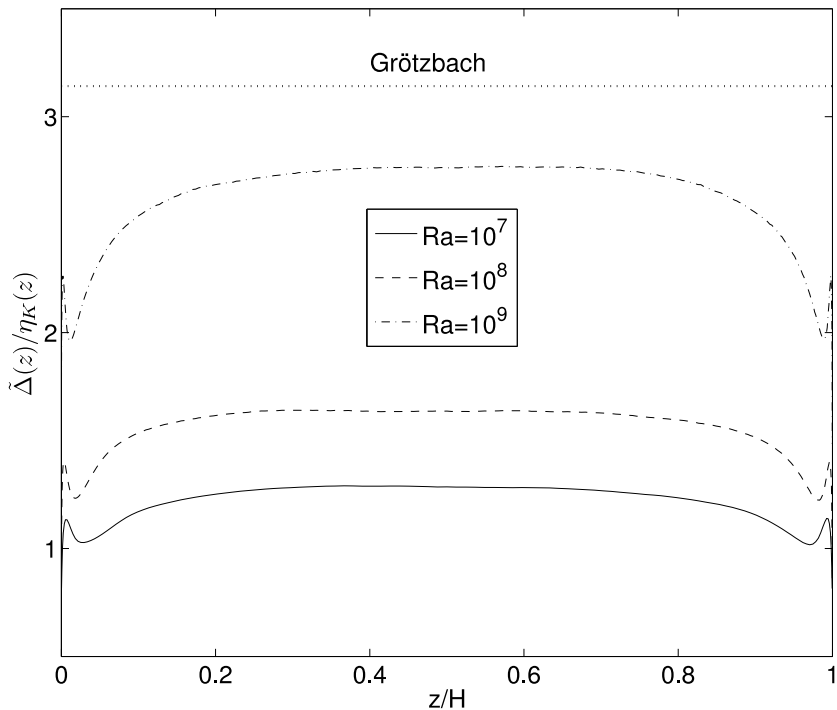}
\caption{Ratio of the maximum geometric mean grid spacing $\tilde{\Delta}(z)$ and the 
Kolmogorov scale $\eta_K(z)$ for three different Rayleigh numbers in cells with $\Gamma=1$. 
The horizontal dotted line indicates the global Gr\"otzbach resolution criterion $\Delta/
\eta_K\le \pi$ (Gr\"otzbach 1983) where $\Delta$ is the global geometric mean grid spacing.} 
\label{fig1}
\end{center}
\end{figure}

The equations are discretized on a staggered grid with a second-order finite 
difference scheme (Verzicco \& Orlandi 1996; Verzicco \& Camussi 2003). The pressure field $p$ 
is determined by a two-dimensional Poisson solver after applying a one-dimensional Fast Fourier 
Transform (FFT) in the azimuthal direction. The time advancement is done by a third-order 
Runge-Kutta scheme. The grid spacings are non-equidistant in the radial and vertical directions. 
In the vertical direction, the grid spacing corresponds to the Tschebycheff collocation points. 
The grid resolutions employed in all runs are listed in Table 1.  The numerical effort grows 
with $\Gamma^2$ in the horizontal circular plane since the resolution at the sidewalls has to be maintained. 
The smallest mean scale in a turbulent flow is the Kolmogorov dissipation length which is usually 
defined as
\begin{equation}
\eta_K=\frac{\nu^{3/4}}{\langle\epsilon\rangle^{1/4}}\,,
\end{equation}
where $\langle\epsilon\rangle$ is the mean of the energy dissipation rate (see e.g. Pope 2000), which is given by
\begin{equation}
\epsilon({\bm x},t)=\frac{\nu}{2} \left(\frac{\partial u_i}{\partial x_j}+\frac{\partial u_j}{\partial x_i}\right)^2.
\end{equation}
The symbol $\langle\cdot\rangle$ stands for a statistical average. The resolution criteria based on 
$\eta_K$ works well in homogeneous isotropic turbulence, but has to be modified for the inhomogeneous 
situation.  We define a height-dependent Kolmogorov scale as
\begin{equation}
\eta_K(z)=\frac{\nu^{3/4}}{\langle\epsilon(z)\rangle_{A,t}^{1/4}}\,.
\end{equation}
The symbol $\langle\cdot\rangle_{A,t}$ denotes an average over a plane at a fixed height $z$ 
and an ensemble of statistically independent snapshots.  Following Emran \& Schumacher (2008), 
we define the maximum of the geometric mean of the grid spacing at height $z$ by
$\tilde{\Delta}(z)=\max_{\phi,r}[\sqrt[3]{\Delta_{\phi}(z)\Delta_r(z)\Delta_z(z)}]$. Fig.~\ref{fig1} 
plots the ratios $\tilde{\Delta}(z)/\eta_K(z)$ over the cell height for three different Rayleigh numbers. 
One can observe that the ratio varies close to the upper and lower plates and levels off in the bulk. 
Overall, it does not exceed the global resolution criterion by Gr\"otzbach (1983), $\Delta/\eta_K
\le \pi$, for the given Rayleigh numbers.  

\begin{table}
\begin{center}
\begin{tabular}{rccccc}
$N_{\phi}\times N_r\times N_z$ & $Ra$ & $\Gamma$ & $t/t_f$ & $Nu\pm\sigma$ & 
$\sigma\;\;\mbox{in}\;\;\%$  \\
& & & & & \\
 $97\times65\times128$ & $10^7$ & 0.50 & 300 & 17.08$\pm$0.07 & 0.4\\
$193\times97\times128$  & $10^7$ & 1.00  & 150 & 16.73$\pm$0.08 & 0.5\\
$217\times133\times128$  & $10^7$ & 1.50  &111  & 16.37$\pm$0.08& 0.5 \\
$217\times133\times128$  & $10^7$ & 1.75  &151  & 16.11$\pm$0.03& 0.2 \\
$217\times133\times128$  & $10^7$ & 2.00  & 250 & 15.88$\pm$0.07 &0.4 \\
$217\times133\times128$ & $10^7$ & 2.25   & 251 & 15.97$\pm$0.04 &0.2 \\
$257\times165\times128$ & $10^7$ & 2.50   & 251 & 15.77$\pm$0.03 & 0.1  \\
$257\times165\times128$ & $10^7$ & 2.75  & 251 & 15.97$\pm$0.04 & 0.3  \\
$257\times165\times128$  & $10^7$ & 3.00 & 150 & 16.06$\pm$0.05 &0.3 \\
$301\times211\times128$  & $10^7$ & 4.00  & 150 & 16.22$\pm$0.03 &0.2 \\
$385\times281\times128$  & $10^7$ & 6.00  & 150 & 16.66$\pm$0.04 &0.2\\ 
$401\times311\times128$  & $10^7$ & 8.00  & 150 &  17.44$\pm$0.02 &0.1\\
$513\times361\times128$  & $10^7$ & 10.00  & 150 &  17.34$\pm$0.03 &0.2\\
$601\times401\times128$  & $10^7$ & 12.00  & 150 &  17.49$\pm$0.03 &0.2\\
& & & & & \\
$151\times81\times160$   & $5\times 10^7$ & 0.50 & 150 & 26.20$\pm$0.21 & 0.8\\
$257\times129\times160$   & $5\times 10^7$ & 1.00    &  150  & 25.86$\pm$0.13 & 0.5\\
$271\times151\times160$  & $5\times 10^7$ & 2.00    &  149   & 25.83$\pm$0.12 & 0.5 \\
$401\times225\times160$  & $5\times 10^7$  & 3.00 &  145   & 25.90$\pm$0.05 & 0.2 \\
& & & & & \\
$151\times101\times256$  & $10^8$ & 0.50 & 300 &  32.06$\pm$0.24 & 0.7\\
$271\times151\times256$  & $10^8$ & 1.00    & 150 &  32.21$\pm$0.32 & 1.0\\
$271\times151\times256$ & $10^8$ & 1.25     & 150 & 31.77$\pm$0.15 & 0.5 \\
$321\times161\times256$  & $10^8$ & 1.50    & 150 &  31.39$\pm$0.11 & 0.3\\
$321\times161\times256$  & $10^8$ & 1.75    & 249 &   31.57$\pm$0.10 & 0.3\\
$361\times181\times256$  & $10^8$ & 2.00    & 145 &  31.25$\pm$0.31 & 1.0\\
$401\times201\times256$ & $10^8$ & 2.25    & 143 &  31.25$\pm$0.21 & 0.7 \\
$401\times201\times256$ & $10^8$ & 2.50    & 146 &  31.87$\pm$0.18 & 0.6 \\
$401\times201\times256$ & $10^8$ & 2.75    & 145 &  32.34$\pm$0.08 & 0.3 \\
$451\times225\times256$  & $10^8$ & 3.00    & 141 &  32.29$\pm$0.12 & 0.4\\
$541\times257\times256$  & $10^8$ & 4.00    & 132 &  33.20$\pm$0.08 & 0.2\\
$801\times451\times256$  & $10^8$ & 8.00    & 81  &  34.78$\pm$0.13 & 0.4\\
& & & & & \\
$201\times101\times310$  & $10^9$ & 0.50 & 150 &  63.67$\pm$0.56 & 0.9\\
$361\times181\times310$  & $10^9$ & 1.00 & 139 &  64.31$\pm$0.64 & 1.0\\
$811\times321\times310$  & $10^9$ & 2.00 & 109 &  63.25$\pm$0.26 & 0.4\\
$1025\times551\times310$  & $10^9$ & 3.00 & 110 &  65.11$\pm$0.50 & 0.8\\
\end{tabular}
\caption{Parameters of simulation runs. The grid resolution, Rayleigh number $Ra$, and 
aspect ratio $\Gamma$ are given. The Prandtl number is $Pr=0.7$ throughout this study. 
Furthermore, the total integration time in units of the free-fall time $t_f=H/U_f$, with 
$U_f=\sqrt{g\alpha \Delta T H}$, and the Nusselt number $Nu$ with standard deviation $\sigma$ 
are given. For all Rayleigh numbers, we find that a characteristic convective velocity 
$U_c=\sqrt{\langle u^2\rangle_{V,t}}$ is about $U_f/5$. Consequently, $t/t_f$ has to be divided
by 5 in order to get $t/t_c$, where the alternative convective time unit is defined as $t_c=H/U_c$. 
This time unit was suggested by van Reeuwijk {\it et al.} (2008) since the standard free-fall 
velocity $U_f$ is too large in comparison with actual turbulent velocity fluctuations.} 
\end{center}
\label{tab1}
\end{table}

\begin{figure}
\begin{center}
\includegraphics[width=13cm]{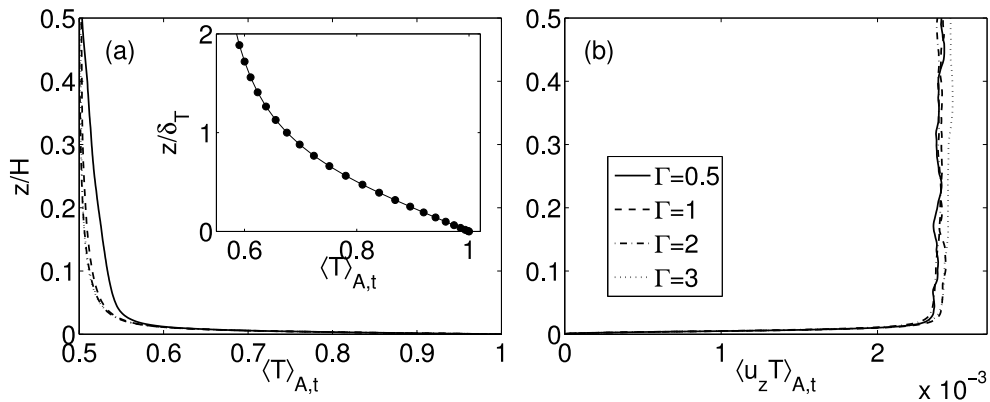}
\caption{The $z$-dependent mean profiles of the (a) temperature  and (b) product of the 
temperature and vertical velocity component. Data for four aspect ratios are plotted at 
Rayleigh number $Ra=10^9$. The inset in (a) magnifies the mean temperature profile 
in the thermal boundary layer for the simulation at $Ra=10^9$ and $\Gamma=3$. Profiles 
are shown for the lower half ($0\le z/H\le 0.5$) of the cell only due to symmetry.} 
\label{fig2}
\end{center}
\end{figure}

Fig.~\ref{fig2} displays the $z$-dependent mean profiles of the temperature and the product of 
the temperature and vertical velocity component. The variations manifested in the profiles contribute 
to the  Nusselt number variation with $\Gamma$ . The inset magnifies the mean temperature 
profile for $Ra=10^9$ and $\Gamma=3$, where the boundary layer is resolved with 17 grid 
planes. 

\section{Dependence of the global heat transfer on aspect ratio} 
\subsection{$Nu(\Gamma)$ at fixed Rayleigh number $Ra$}
The heat transfer through each plane at a fixed height $z$ following the averaging of 
Eq.~(\ref{pseq}) with respect to the horizontal plane is given by
\begin{equation}
Nu(z)=\frac{\langle u_z T\rangle_{A,t}-\kappa \partial_z\langle T\rangle_{A,t}}{\kappa\Delta T/H}=const.
\label{Nup}
\end{equation} 
The global Nusselt number, $Nu$, can then be written as
\begin{equation}
Nu=\frac{1}{H}\int_0^H Nu(z) \mbox{d}z=1+\frac{H}{\kappa\Delta T}\langle u_z T\rangle_{V,t}\,,
\label{Nug}
\end{equation} 
where $\langle\cdot\rangle_{V,t}$ denotes an average over the whole cell volume and an ensemble of 
statistically independent snapshots. The samples are gathered over the total integration time, which
is also listed in Table 1, and given in units of the free-fall time $t_f=H/U_f$, with the free-fall velocity
$U_f=\sqrt{g\alpha\Delta T H}$ as the characteristic velocity. The last columns of the table display
the Nusselt number $Nu$ and the standard deviation $\sigma$, which is calculated as
\begin{equation}
\sigma=\sqrt{\frac{1}{N_z}\sum^{N_z}_{j=1}[Nu(z_j)-Nu]^2}\,.
\end{equation}
Here $z_j$ is the vertical coordinate of each gridplane and $Nu(z)$ and $Nu$ follow from 
Eqns.~(\ref{Nup}) and (\ref{Nug}), respectively. These standard deviations are smaller than or 
equal to 1\% and thus comparable with Kerr (1996). The total integration time of the simulations
is comparable with van Reeuwijk {\it et al.} (2008).  

\begin{figure}
\begin{center}
\includegraphics[width=11cm]{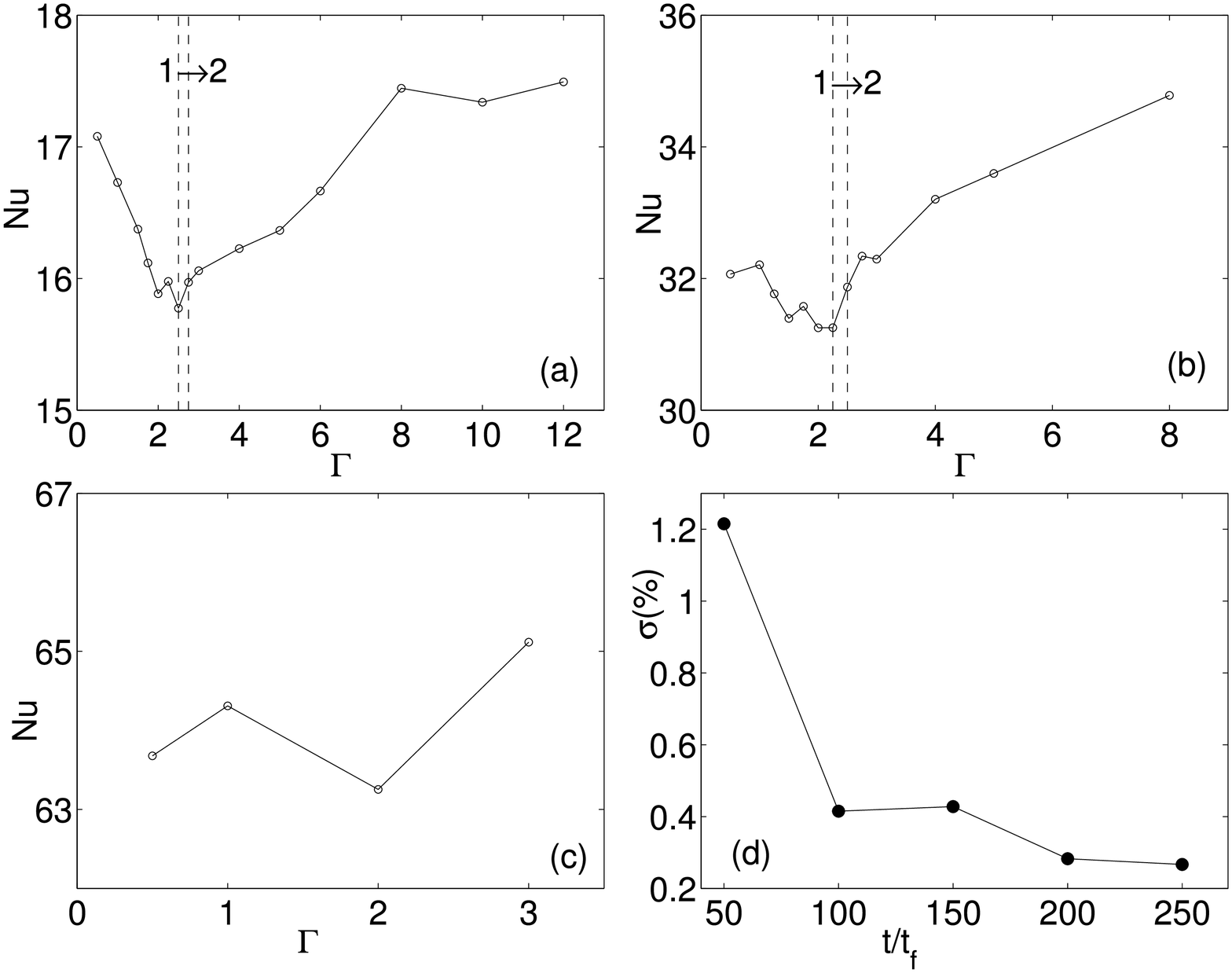}
\caption{Nusselt number $Nu$ as a function of the aspect ratio $\Gamma$ for (a)
$Ra=10^7$, (b) $Ra=10^8$, (c) $Ra=10^9$. The crossover from one circulation roll
to two rolls is indicated in (a) and (b) by two parallel dashed lines. The errorbars
for all Nusselt numbers shown are smaller than the size of the symbols. For each single
snapshot, the Nusselt number was determined as a volume average. In addition, an arithmetic mean is
taken over $N_{samp}=t/t_f$ statistically independent turbulent samples (see Eq.~(\ref{Nug}) and Table 1).
The convergence of the standard deviation $\sigma$ with increasing number of samples 
$N_{samp}=t/t_f$ is shown in (d) for the data set $Ra=10^7$ and $\Gamma=2$.} 
\label{fig3}
\end{center}
\end{figure}

Fig.~\ref{fig3} shows the Nusselt number 
$Nu$ as a function of the aspect ratio, $\Gamma$, for three different Rayleigh numbers, 
namely $Ra=10^7, 10^8$ and $10^9$. At $Ra=10^7$ (Fig.~\ref{fig3}(a)), $Nu$ decreases with 
increasing  $\Gamma$, attains a minimum value at $\Gamma\approx 2.5$, then increases to a 
maximum value close to $\Gamma \approx 8$, and finally saturates for $\Gamma>8$. Variations
of $Nu(\Gamma)$ can also be observed in Figs.~\ref{fig3}(b) and \ref{fig3}(c) for the other two larger 
Rayleigh numbers. The minimum of $Nu(\Gamma)$ is detected at $\Gamma\approx 2.5$ and $\Gamma\approx 2.25$ for $Ra=10^7$  and  $Ra=10^8$, respectively. This is the point where a transition in the LSC from a single-roll to a double-roll pattern will occur (see section 4). 
On the basis of stability analysis, Oresta {\it et al.} (2007) have shown that there is always a single-roll for $\Gamma\leq 2$ in the weakly nonlinear regime irrespective of the initial conditions. However, our Rayleigh numbers here are in fully turbulent regime. With our present computing capability, we could not go beyond $\Gamma> 8$ 
for $Ra=10^8$ and $\Gamma>3$ for $Ra=10^9$. In particular, for the largest Rayleigh number, 
we can provide four data points only and, therefore, the minimum of $Nu(\Gamma)$ is inconclusive in this case, although it is apparently at $\Gamma\approx 2$ in Fig.~\ref{fig3}(c). On the basis of our simulation data, we can not conclude exactly at which aspect ratio the Nusselt numbers become independent of the cell geometry for all the Rayleigh numbers, however, the trend indicates that it is at $\Gamma\approx 8$ for $Ra=10^7$ and  $\Gamma\gtrsim 8$ for $Ra=10^8$. The variations in $Nu$, as defined by 
the difference between the maximum and minimum in the Nusselt number series, are significant -- especially for the lower Rayleigh numbers -- and yield 10.9\%, 11.3\%, and 3.0\% for $Ra=10^7$, $10^8$ and $10^9$ respectively. 

A closer inspection of the three panels in Fig.~\ref{fig3} reveals non-monotonic graphs of 
$Nu(\Gamma)$ with local maxima and minima, in particular for the two larger Rayleigh numbers. 
We have first verified that there is sufficient statistical convergence of the data (see Table 1). Since statistical 
uncertainties can be excluded, there must be physical reasons for the behaviour observed in Fig.~\ref{fig3}. We observe that the time-averaged flow patterns in the turbulent cell are similar to those at the onset of convection (Figs.~\ref{fig5} and 6). In this case, an integer number of rolls 
must fit into the cell. This is exactly the reason why, for example, the linear instability studies by 
Koschmieder (1969) and Charlson \& Sani (1970, 1971) in the cylindrical cells with insulated side-walls 
yield stability curves $Ra_{cr}(\Gamma)$ with local extrema in the low--$\Gamma$ regime, and extend 
to an asymptotic value for larger $\Gamma$ only. Small discontinuities in $Nu(Ra)$ in 
the weakly nonlinear regime, which could be traced back to a change of the number of rolls in the cell, 
have been also reported by Gao  {\it et al.} (1987). 

These pattern bifurcations can be studied when a small number of degrees of freedom 
dominates the dynamics. It is not obvious that in a fully turbulent case, where infinitely 
many degrees of freedom exist,  coherent patterns exist and prevail. Similar patterns 
can, however, be found in a turbulent Taylor vortex flow at high Reynolds number 
(Lathrop {\it et al.} 1992). The POD analysis in section 5 demonstrates  that the 
LSC carries a significant amount of heat through the cell. We also show that a 
change of the LSC morphology causes jumps in the amount of heat transported by the first few POD modes. These findings strengthen our observation of $\Gamma$-dependent  heat transfer (see Fig.~\ref{fig3}). It should also be mentioned that persistent coherent patterns at larger Rayleigh numbers have been emphasized 
by Busse (2003) as a sequence-of-bifurcations to the turbulent state.
 
\subsection{$Nu(Ra)$ at fixed aspect ratio $\Gamma$}
Systematic experiments with various values of  $\Gamma$ larger than unity were conducted  
by three groups. First, Wu \& Libchaber (1992) detected a power law scaling with $Ra$, namely
\begin{equation}
Nu(Ra,\Gamma)=A(\Gamma)\times Ra^{\beta}\,.
\label{Nu1}
\end{equation}      
Their measurements indicated almost an unchanged exponent $\beta$ and an 
aspect-ratio-dependent prefactor. Second, Sun {\it et al.} (2005) suggested the following 
scaling law on the basis of their experiments as
\begin{equation}
Nu(Ra,\Gamma)=A_1(\Gamma)\times Ra^{\beta_1}+A_2(\Gamma)
\times Ra^{\beta_2}\,.
\label{Nu2}
\end{equation}      
This scaling is a combination of two power laws with $\beta_1=1/3$ and $\beta_2=1/5$. Again, 
the prefactors depend on $\Gamma$ and a saturation of the Nusselt number $Nu$ for $\Gamma\ge 10$ 
has been detected. Third, Funfschilling {\it et al.} (2005) did not observe any sensitivity of the heat transfer on the aspect ratio. Their measurements gave power laws of the form 
\begin{equation}
Nu(Ra,\Gamma)=A\times Ra^{\beta}\,,
\label{Nu3}
\end{equation}      
but with a continuous drift of the exponent from $\beta=0.28$ at $Ra\sim 10^8$ up to $\beta=0.33$ 
at $Ra\gtrsim 10^{10}$.  Their results were essentially unaltered by an increase in the aspect ratio.
On the numerical side, a power law of $Nu\sim\Gamma^{-1}$ for $\Gamma\leq3$ was obtained by 
Ching \& Tam (2006) on the basis of two-dimensional steady state calculations. 

The present data allows us to compare our results with the scaling laws given in (\ref{Nu1})--(\ref{Nu3}). 
Table~2 displays the fit results for power laws in the form  $Nu=A\times Ra^{\beta}$ at fixed 
aspect ratios $\Gamma=1/2, 1, 2$ and 3.  Each data series contains four Rayleigh numbers, namely 
$Ra=10^7$, $5\times10^7$, $10^8$ and $10^9$. Within this range of $Ra$, we observe a growth 
of the  exponent $\beta$ from 0.287 to 0.305, which is about 6\% variation. The present scaling law 
for $\Gamma=1$ differs slightly from the earlier reported scaling of $Nu=0.175\times Ra^{0.283}$ 
in Emran \& Schumacher (2008). In the former case, six Rayleigh numbers from $5\times 10^6$ to 
$10^9$, but fewer snapshots for the higher Rayleigh numbers, were included. This demonstrates the 
sensitivity of the scaling laws and demands additional efforts to be taken here.  Both the prefactor $A$ 
and exponent $\beta$ seem to be functions of the aspect ratio and the functional form is thus  
\begin{equation}
Nu(Ra,\Gamma)=A(\Gamma)\times Ra^{\beta(\Gamma)}\,.
\label{Nu4}
\end{equation}      
\begin{figure}
\begin{center}
\includegraphics[width=13cm]{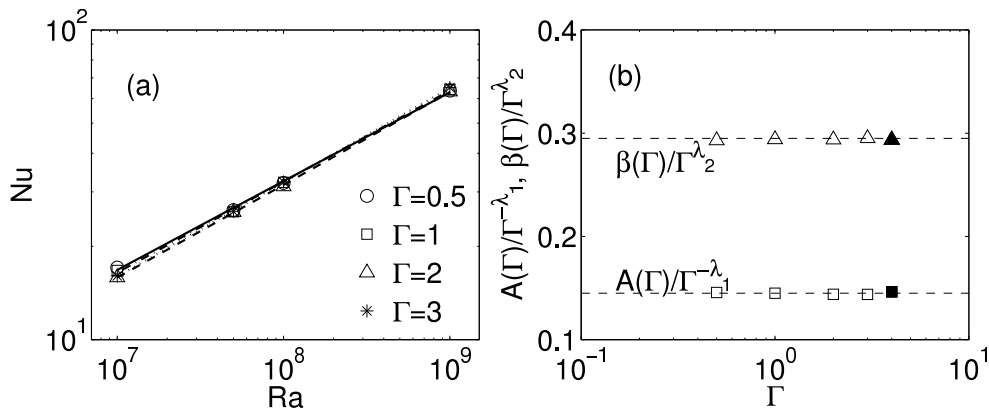}
\caption{Aspect ratio dependence of the fit coefficients, $A(\Gamma)$ and $\beta(\Gamma)$. (a): Data
for the Nusselt number and power law fits to $Nu(Ra)$ as reported in Table 2. (b): Compensated power law plots
for $A(\Gamma)$ and $\beta(\Gamma)$. The exponents are $\lambda_1=0.18$ for $A$ and $\lambda_2=0.03$ 
for $\beta$. The open symbols are the present simulation data. The filled symbols correspond to 
Niemela \& Sreenivasan (2006). We have fitted their data from $Ra=1.10\times 10^8$ to 
$9.51\times 10^9$ (see their Table 1). } 
\label{fig4}
\end{center}
\end{figure}
Fig.~\ref{fig4}(a) shows power law fits (\ref{Nu4}) to our DNS data for several aspect ratios and Fig.~\ref{fig4}(b) shows $A(\Gamma)\sim \Gamma^{-\lambda_1}$ and $\beta(\Gamma)\sim\Gamma^{\lambda_2}$ in a 
compensated form for $0.5\le \Gamma\le 3$. 
The measurements that come closest to the present study, both in Rayleigh  and Prandtl 
numbers,  are those by Niemela \& Sreenivasan (2006) at $\Gamma=4$. A power law fit of their 
data  for $1.10\times 10^8\le Ra\le 9.51\times 10^{9}$ yields $Nu=0.114\times Ra^{0.306}$. 
Adding these parameters to Fig.~\ref{fig4} covers data over almost a decade of $\Gamma$. 
We see that both parameters, $A$ and $\beta$, almost perfectly follow the power law with respect to $\Gamma$. The 
exponent for $\beta$ is $\lambda_2=0.03$, which is small. The dependence of the prefactor $A$ on $\Gamma$ is 
stronger, with $\lambda_1=0.18$.  It is clear that further studies are required to determine whether this weak dependence on 
$\Gamma$ prevails at larger Rayleigh numbers or not. Furthermore, we can expect that, for sufficiently 
large $\Gamma$, both exponents will saturate to aspect-ratio-independent values. This was shown clearly in
Fig.~\ref{fig3} for $Ra=10^7$.  In addition,  the saturation threshold for $A$ and $\beta$ most likely depends on the Prandtl number, 
which is constant in our case. 

\renewcommand{\arraystretch}{1.4}
\begin{table}
\begin{center}
\begin{tabular}{ccccc}
\multicolumn{1}{l}             
\mbox{Fit Coefficients}    &  $\Gamma=\frac{1}{2}$     & $\Gamma=1$     & $\Gamma=2$     & $\Gamma=3$   \\               
$A$       & $0.165$ &  $0.145$ & $0.127$& $0.118$\\
$\beta$ & $0.287$ &  $0.294$ & $0.300$& $0.305$
\end{tabular}
\label{tab2}
\caption{Nusselt number as a function of the Rayleigh number for different aspect ratios.  The scaling
$A\times Ra^{\beta}$ has been fit for four aspect ratios. }
\end{center}
\end{table}
\begin{figure}
\includegraphics[width=4cm]{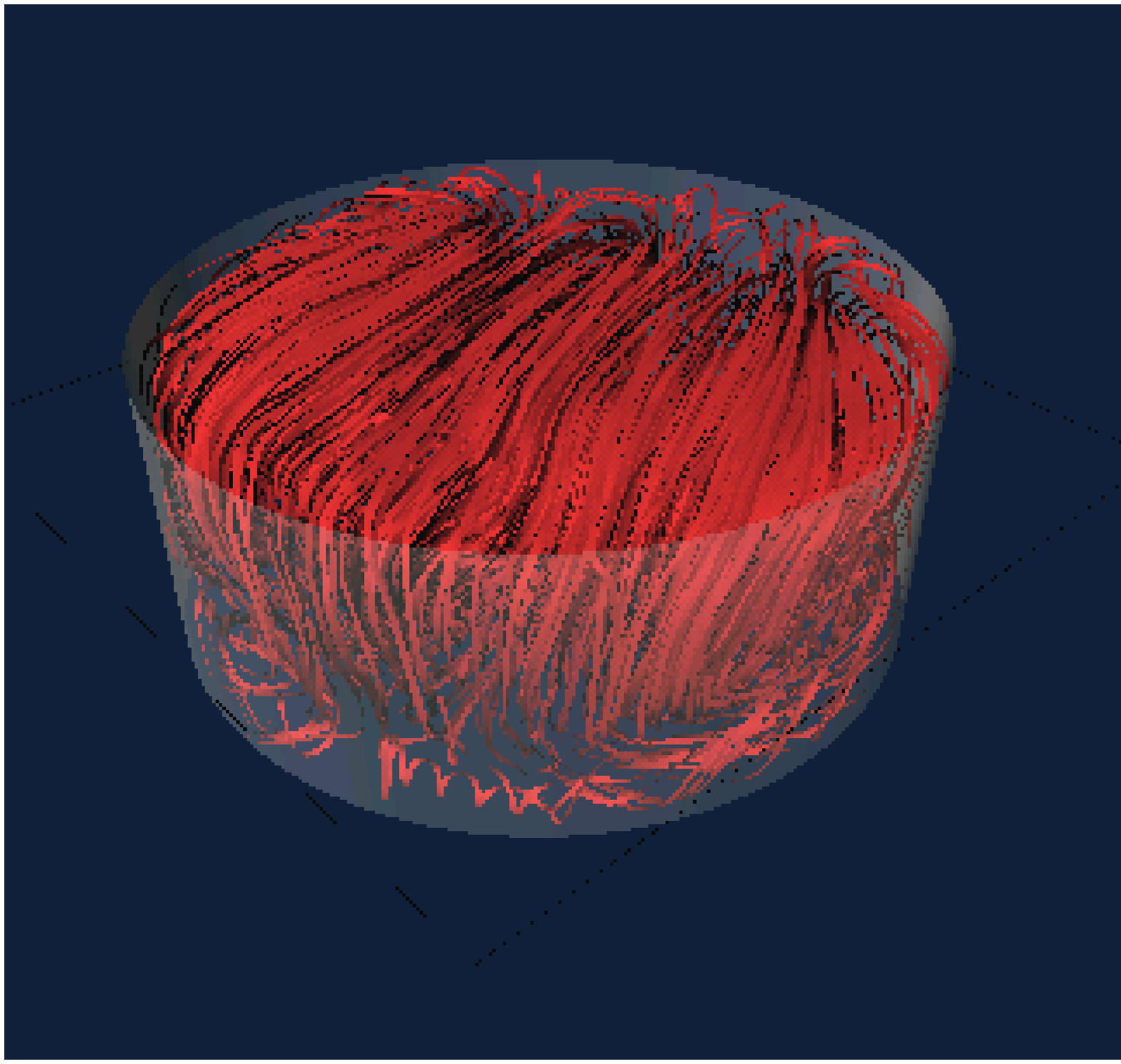}
\hspace{0.2cm}
\includegraphics[width=4cm]{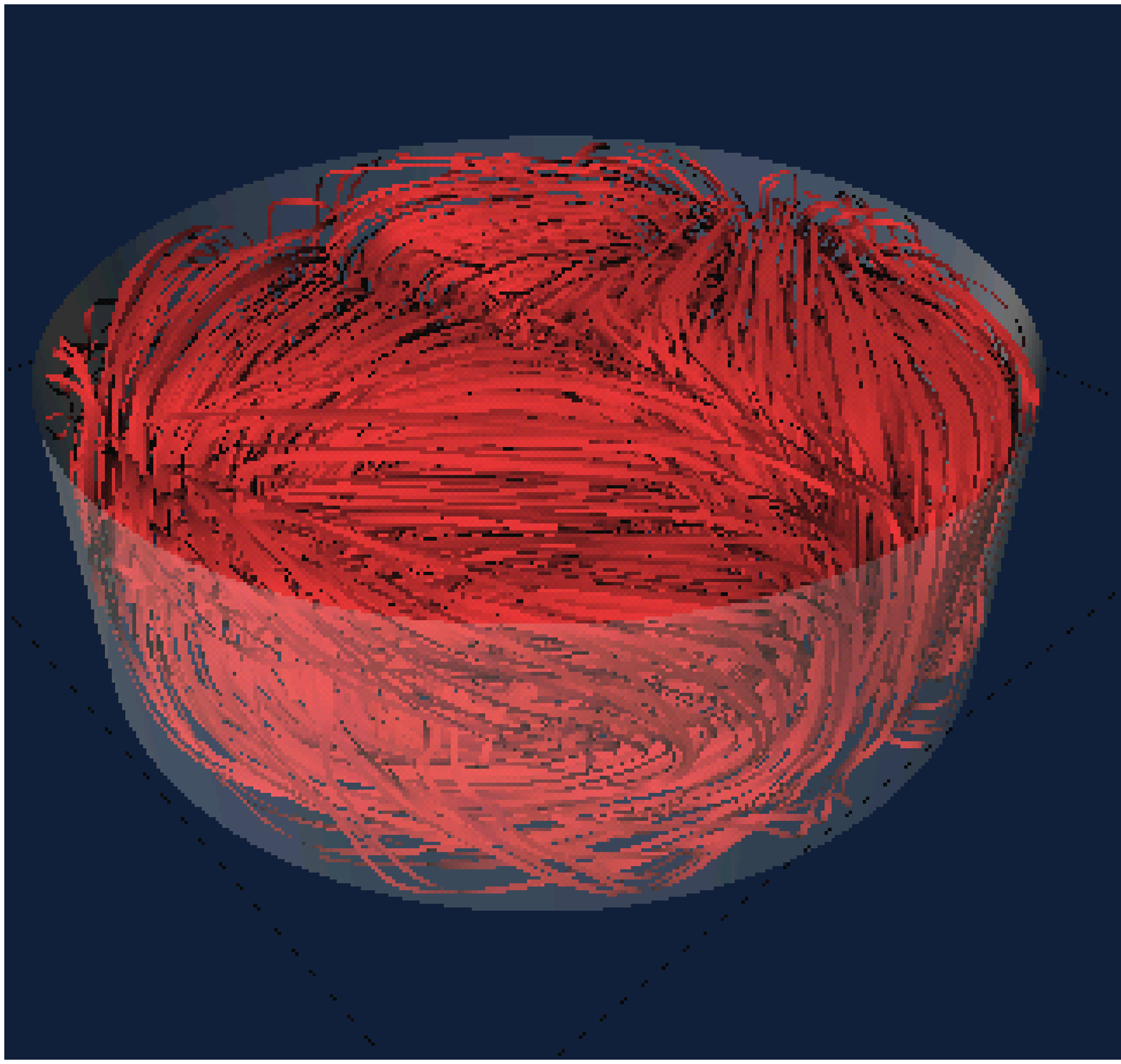}
\hspace{0.2cm}
\includegraphics[width=4cm]{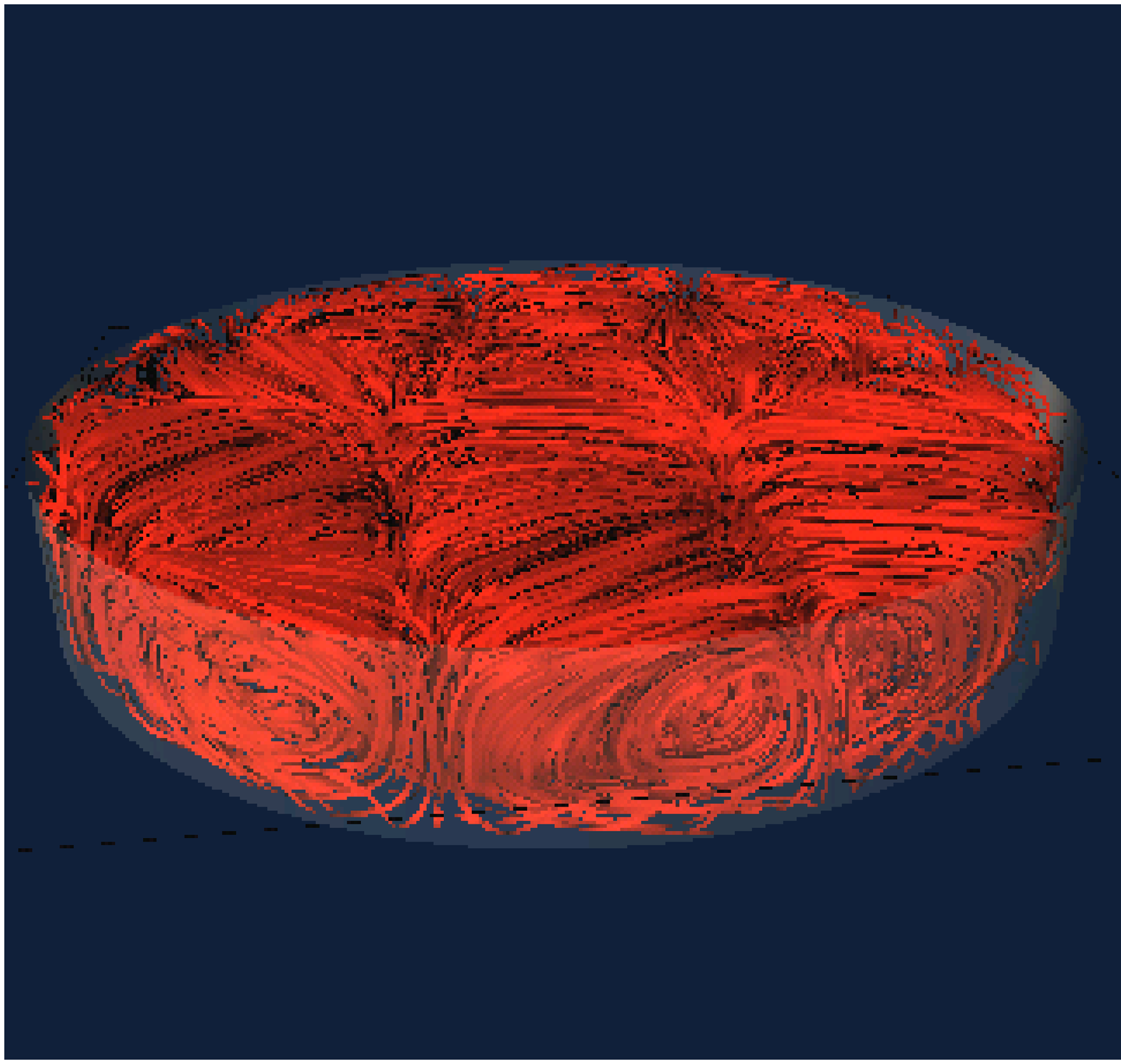}

\vspace{0.2cm}
\includegraphics[width=13cm]{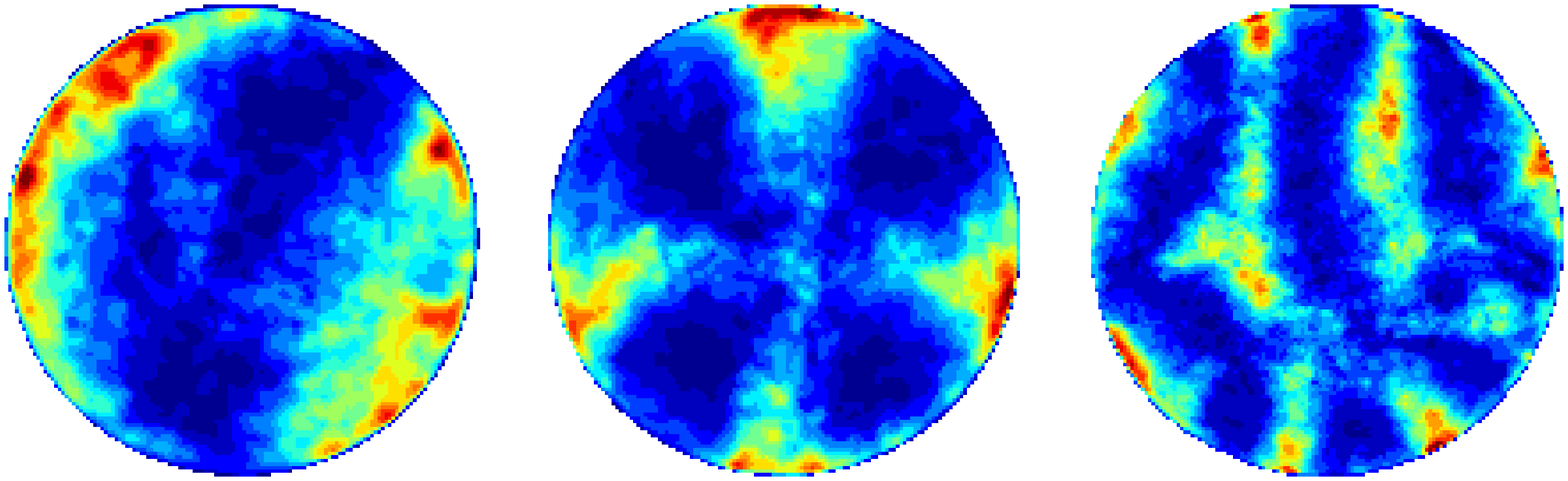}

\caption{Flow patterns at different aspect ratios. 
Streamlines (top row) and contours of the local heat transfer $u_z\theta$ (bottom row) 
for three different aspect ratios, $\Gamma=2.5$ (left column), $\Gamma=3$ (middle column), 
and  $\Gamma=6$ (right column), at $Ra=10^7$ are presented. All data are obtained by time 
averaging a sequence of 50 statistically independent snapshots. In the bottom row 
$\langle u_z \theta\rangle_t(r,\phi,z=1/2)$ is shown.} 
\label{fig5}
\end{figure}
\section{Large-scale circulation}
Let us now investigate the behaviour of the LSC. In Fig.~\ref{fig5}, we present the LSC 
for three aspect ratios $\Gamma=$2.5, 3, and 6 at $Ra=10^7$. The streamline plots in the 
upper three panels have been obtained by averaging the velocity field over 50 consecutive 
snapshots. These snapshots are separated from each other by $\Delta t=t_f=H/U_f$.  
Averaging over three disjoint sequences of 50 snapshots leaves the observed 
LSC patterns unchanged. We conclude, therefore, that the detected LSC pattern is not  
transient. Transient behaviour and large-scale saturation have been investigated by von
Hardenberg {\it et al.} (2008). The time-averaging over the coarse sequence of snapshots
removes not only all small-scale fluctuations of the velocity field, but also oscillations of the 
LSC, which have been observed in 
recent experiments (e.g. Xi \& Xia (2008) and Brown \& Ahlers (2008)), mostly for 
$\Gamma\le 1$. Between $\Gamma=2.5$ and 2.75, the system bifurcates from a one-roll 
to a two-roll pattern. We have also identified this crossover in LSC between $2.25<\Gamma<2.5$ 
for $Ra=10^8$. However, for $Ra=10^9$ we have noticed a single-roll circulation pattern at 
$\Gamma=2$ and a triple-roll pattern at $\Gamma=3$. Here, the LSC patterns for aspect ratios  between 
2 and 3 were not investigated for the highest Rayleigh number. A single-roll at $\Gamma=2$ is consistent with 
the findings of Sun {\it et al.} (2005), Oresta {\it et al.} (2007) and Bukai {\it et al} (2009). The crossovers of the LSC are 
marked in Fig.~\ref{fig3}(a) and Fig.~\ref{fig3}(b) by two  parallel dashed lines. With increasing aspect ratio, 
the LSC becomes a more complex multi-roll configuration, as can be seen in the third column of 
Fig.~\ref{fig5} for $\Gamma=6$. 

In the lower row of Fig.~\ref{fig5}, we show the corresponding contour plots of  $\langle u_z 
\theta\rangle_{t}$ at the midplane where 
\begin{equation}
\theta({\bm x},t)=T({\bm x},t)-\langle T(z)\rangle_{A,t}\,.
\label{turbdata}
\end{equation}
The quantity $u_z\theta$ is the local convective heat flux contribution and $u_z\theta>0$ if rising and falling 
plumes are present. The appearance of rising and falling plumes (red in $\langle u_z \theta\rangle_{t}$ 
contours) in the three panels (lower row of Fig.~\ref{fig5}) is directly correlated to the corresponding 
LSC pattern of the time averaged velocity field. We have also verified that almost the same 
pattern holds for the fluctuations of the local heat transfer, as given by $\langle (u_z \theta)^2\rangle_{t}$.      
\begin{figure}
\begin{center}
\includegraphics[width=6cm]{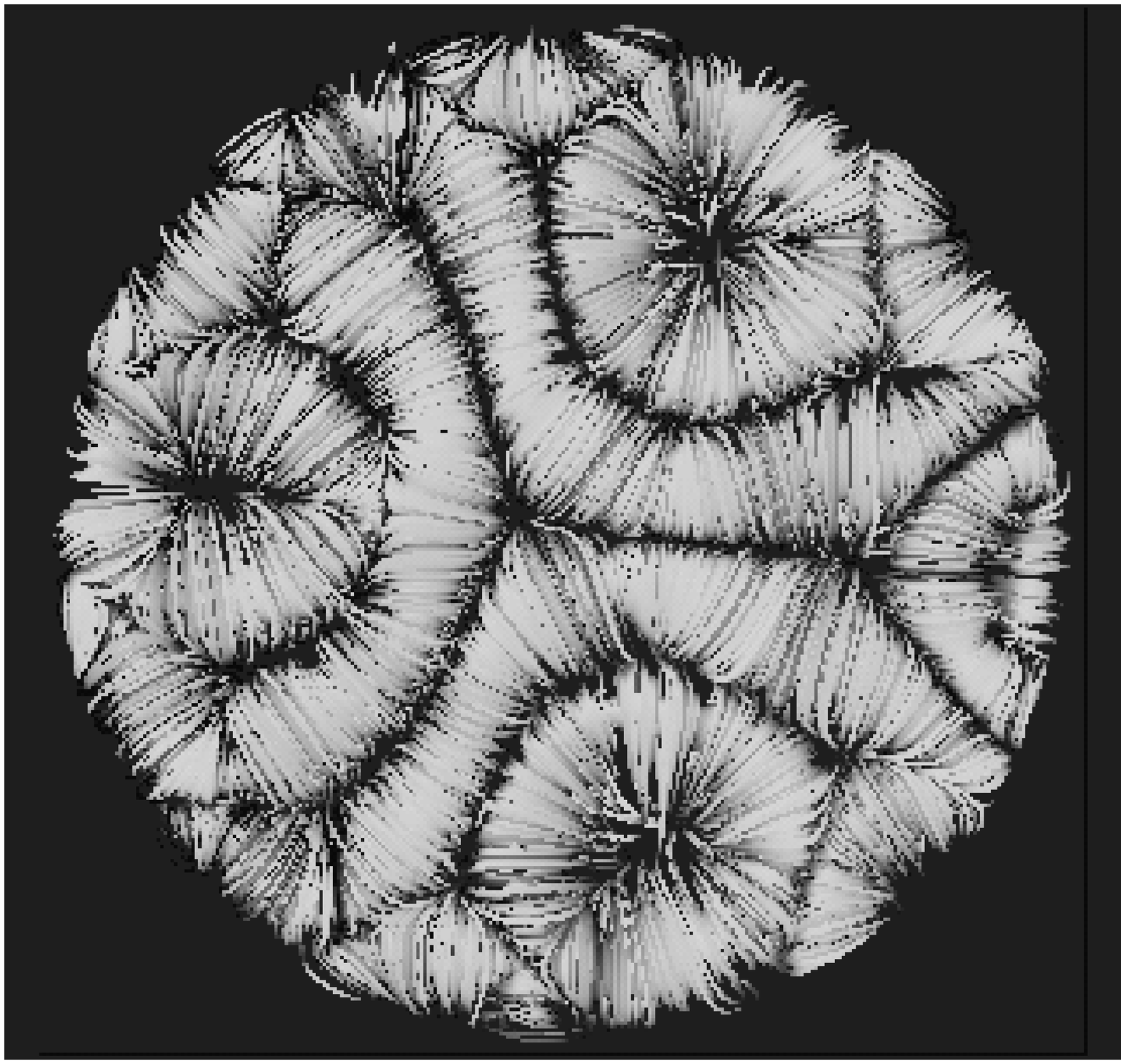}
\hspace{0.2cm}
\includegraphics[width=6cm]{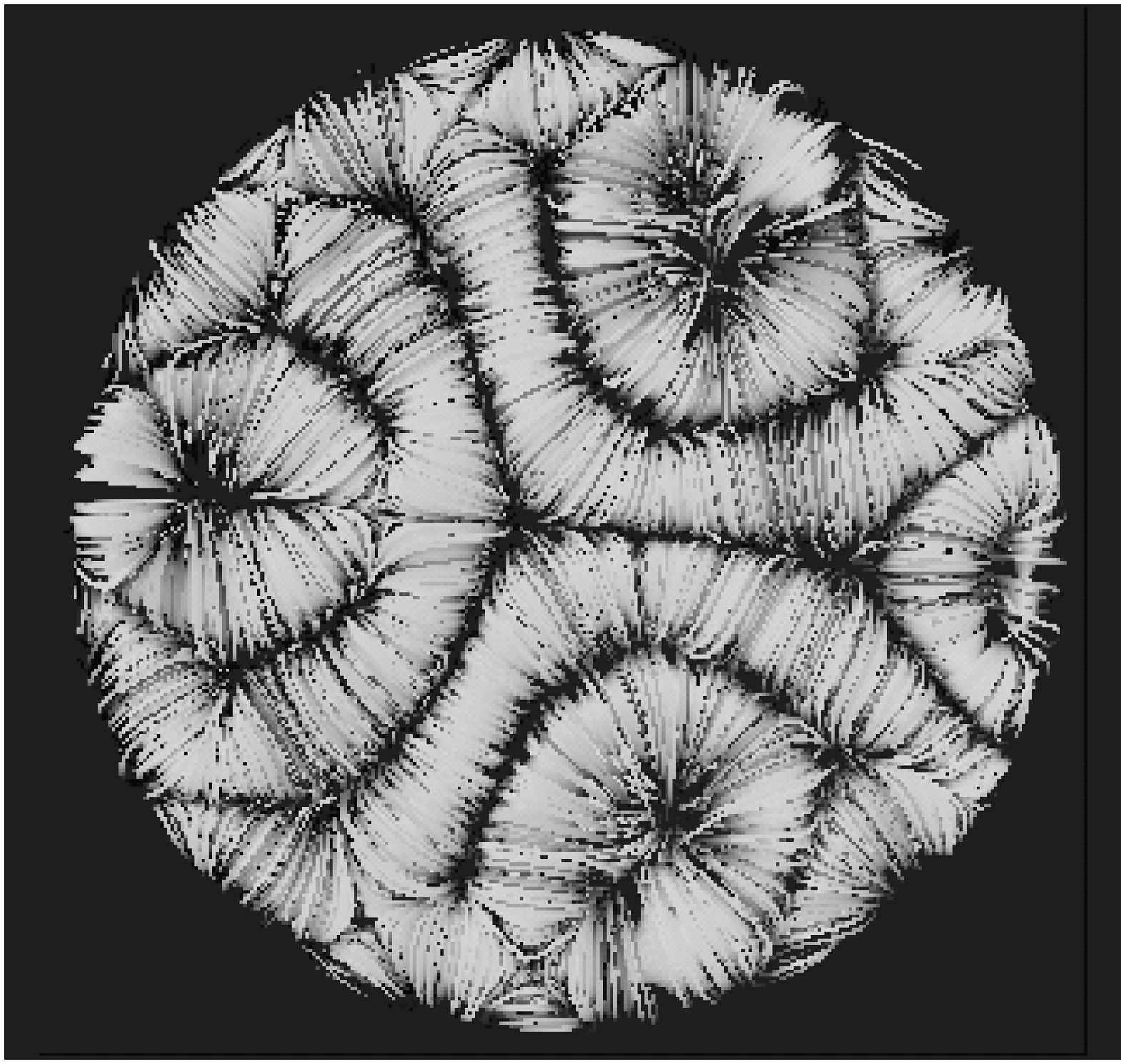}

\vspace{0.2cm}
\includegraphics[width=6cm]{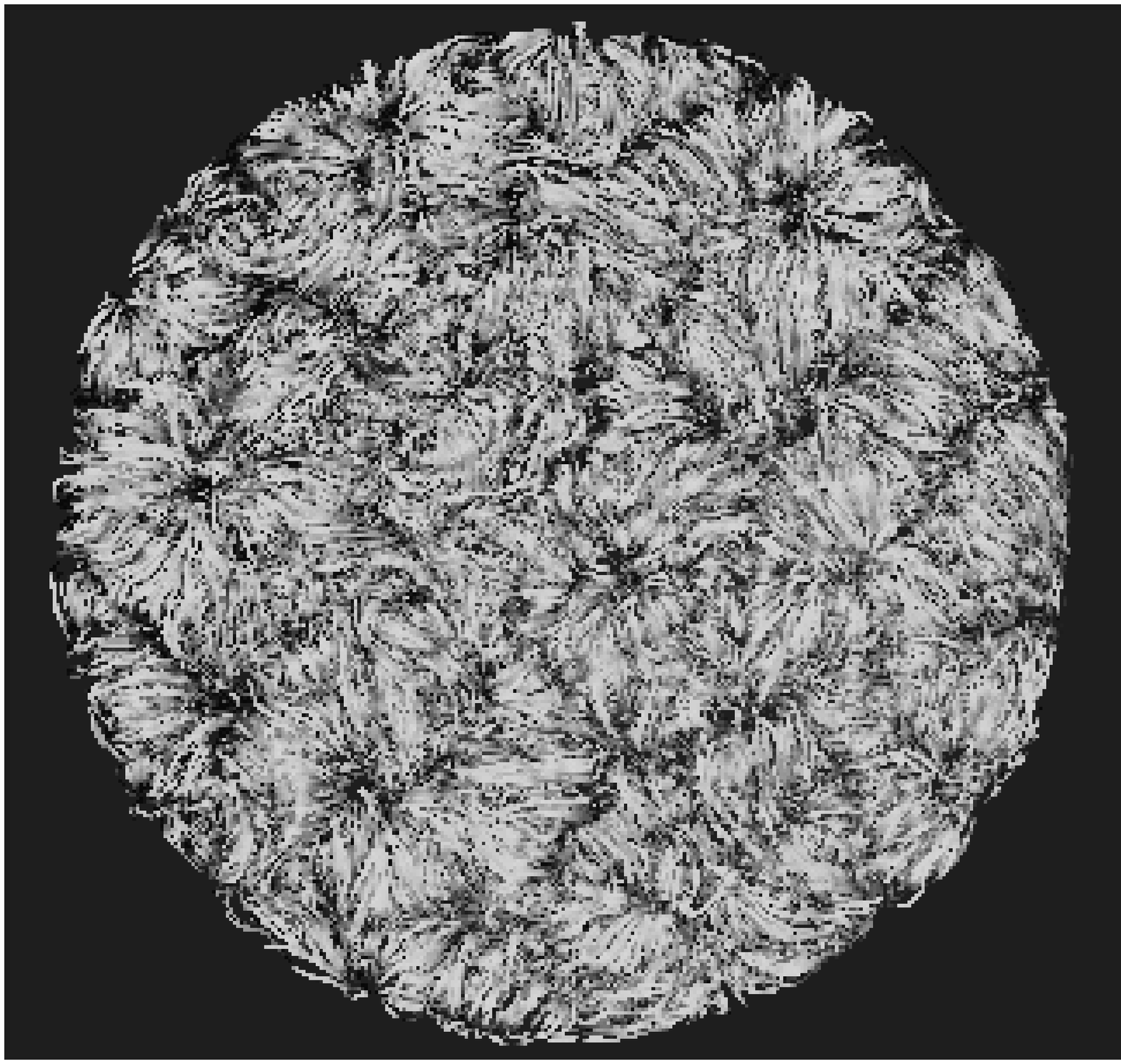}
\hspace{0.2cm}
\includegraphics[width=6cm]{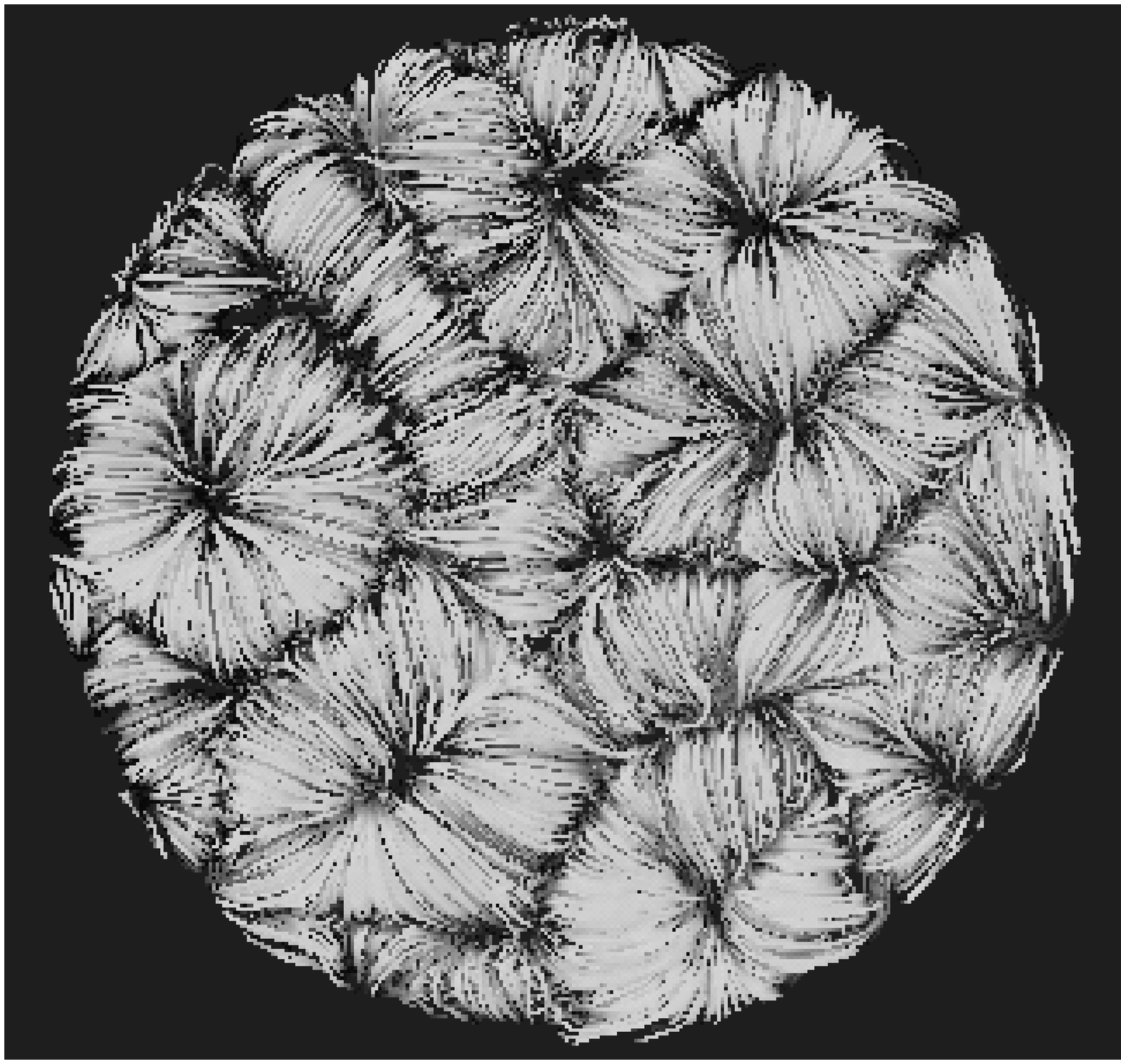}
\caption{Top view of the large scale circulation (LSC) patterns at $\Gamma=12$ for two different Rayleigh 
numbers, $Ra=6\times 10^3$ (top row) and $Ra=10^7$ (bottom row). The streamlines of the instantaneous 
(left column) and time-averaged (right column) velocity field are shown here. In both cases, the time 
averaging is done over 50 convective time units $t_f$.} 
\label{fig6}
\end{center}
\end{figure}

\begin{figure}
\includegraphics[width=4.3cm]{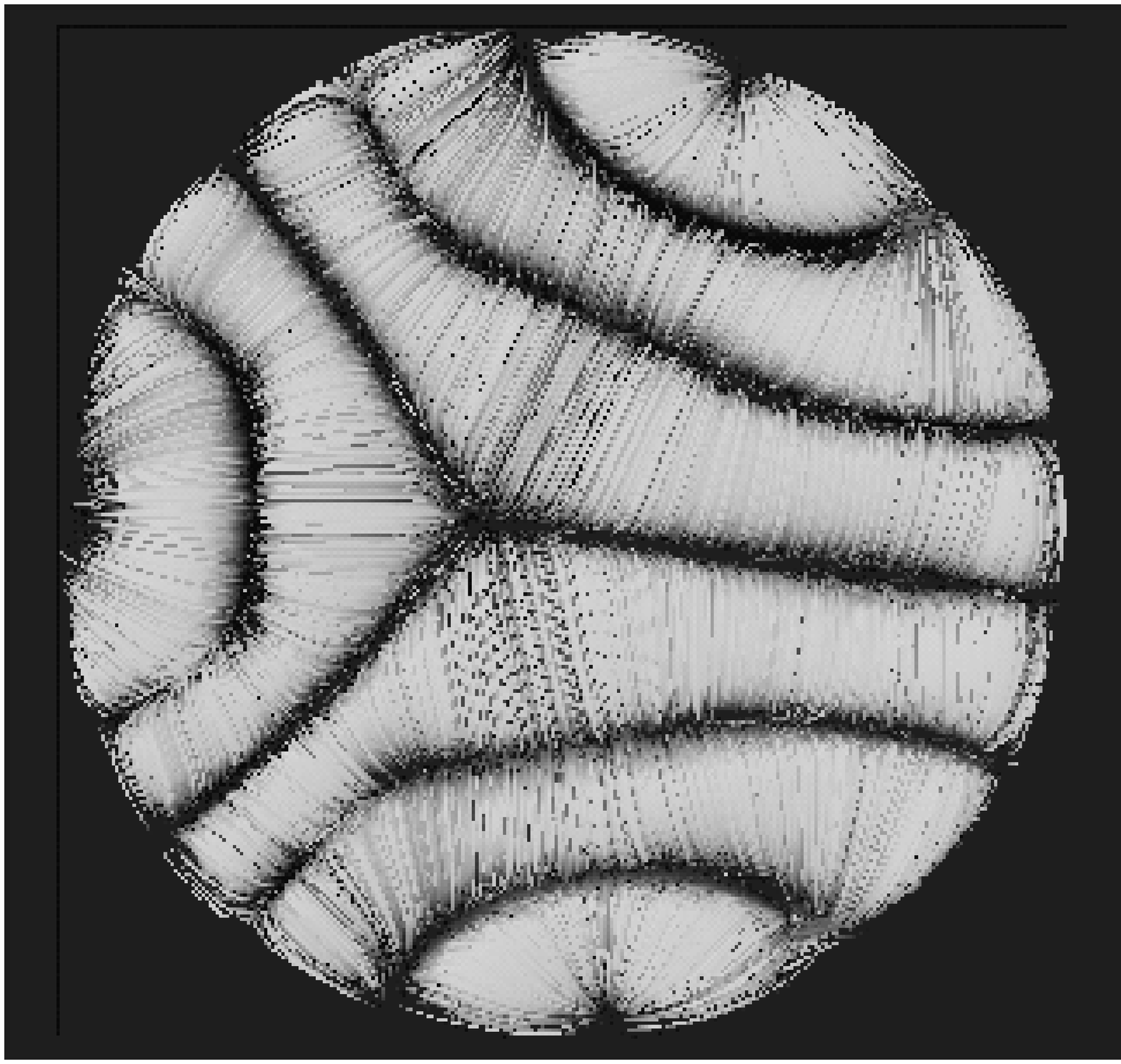}
\hspace{0.1cm}
\includegraphics[width=4.3cm]{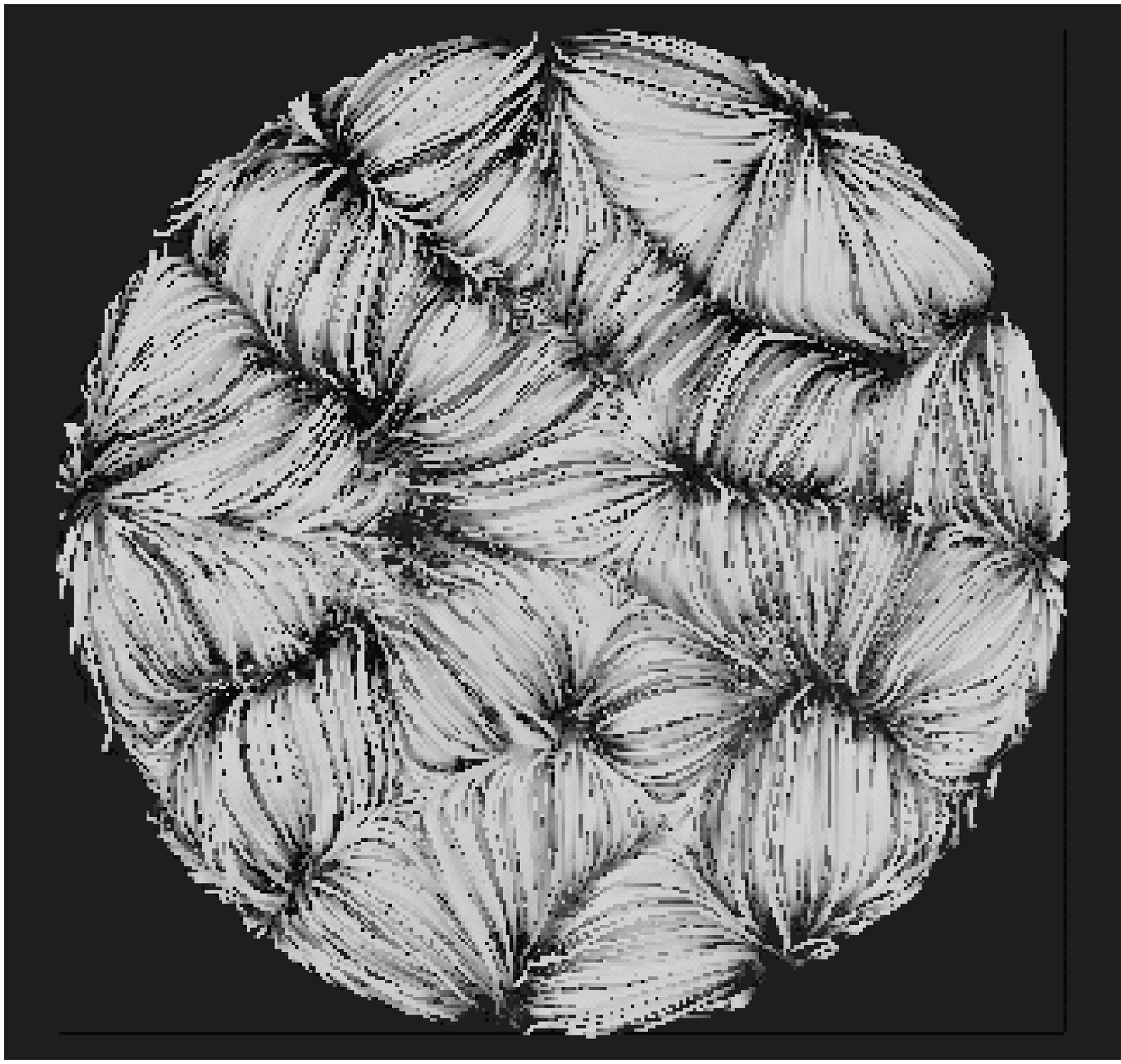}
\hspace{0.1cm}
\includegraphics[width=4.3cm]{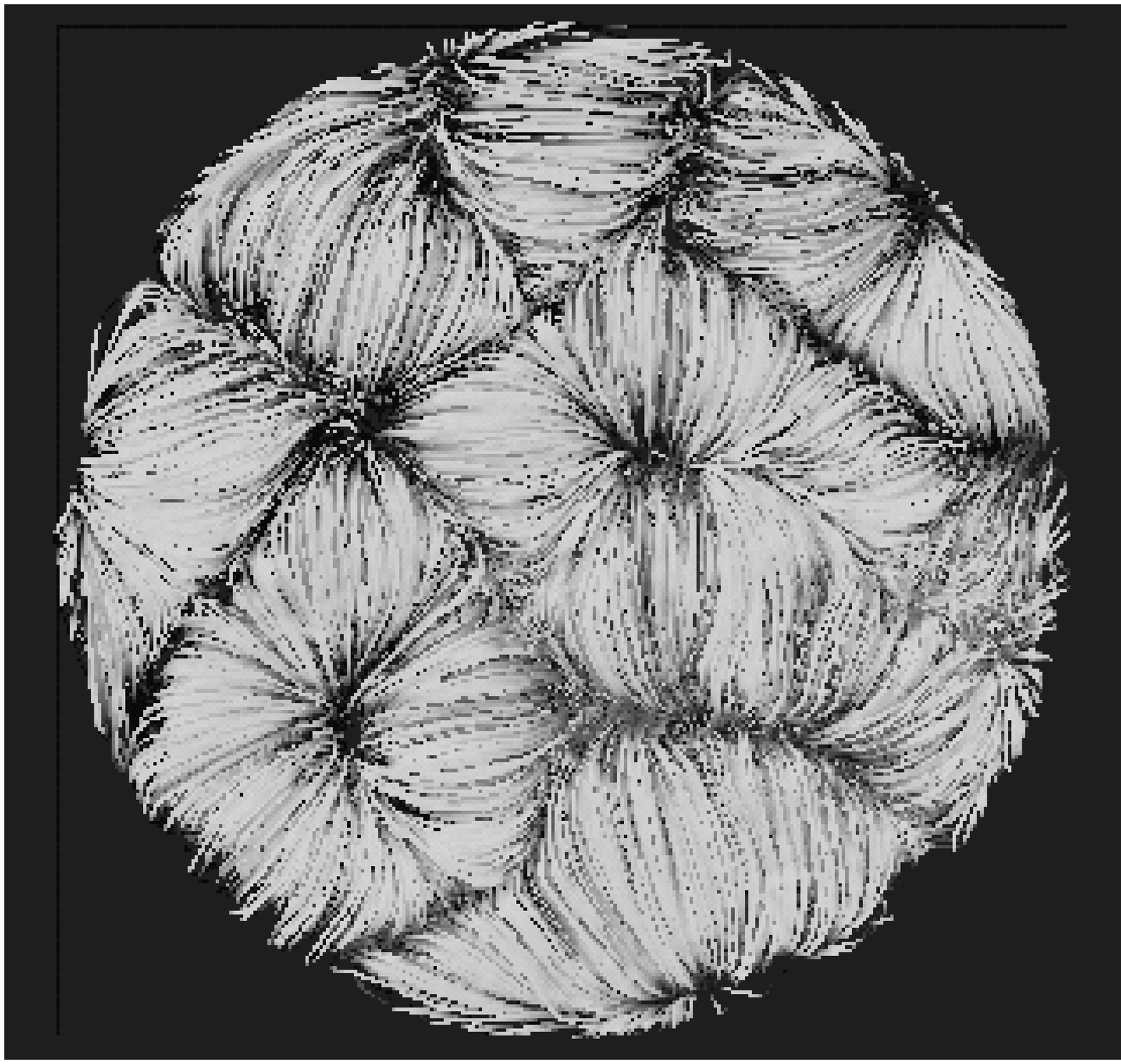}
\caption{Rayleigh number dependence of the large scale circulation. Left panel: 
$Ra=6000$. Mid panel: $Ra=10^7$. Right panel: $Ra=10^8$. All data are for $\Gamma=8$.} 
\label{fig7}
\end{figure}

As already indicated in Fig.~\ref{fig5}, the LSC becomes more complex when the aspect ratio becomes 
larger. For Rayleigh number $10^7$, we were able to run  a numerical simulation up to $\Gamma=12$. 
Fig.~\ref{fig6} reveals such a complex LSC pattern in  convective flow for two different Rayleigh numbers, 
$Ra=10^7$ and $Ra=6000$, at $\Gamma=12$. The left column shows the top view of the streamlines 
for instantaneous snapshots of both simulations, while the right column shows the time-averaged velocity 
field as in Fig.~\ref{fig5}. When the small-scale turbulence (see lower left panel) is filtered out, the resulting 
pattern is strikingly similar to the weakly nonlinear regime right above the onset of convection. We 
observe extended rolls and pentagon-like cells. These patterns have been reported, for example, in 
experiments by Croquette (1989) with argon at $Pr=0.69$ for Rayleigh numbers $Ra\approx 2 Ra_c$, where 
$Ra_c$ is the critical Rayleigh number of the onset of convection. Fig.~\ref{fig7} adds further support to 
the Rayleigh-number-dependence of the LSC.  The left panel nicely displays the extended roll patterns 
in the weakly nonlinear regime at $Ra=6000$ and $\Gamma=8$. Relics of these patterns are still present in the 
turbulent regime at $Ra=10^7$ (mid panel). For 
the largest Rayleigh number, $Ra=10^8$, the LSC is transformed into a pentagon-like cell structure.
Similarly, if we compare the top-right panel of Fig.~\ref{fig6} with the left panel of Fig.~\ref{fig7}, we 
see that there is a reorganization of flow from the roll shape to pentagonal or hexagonal structures with 
increasing $\Gamma$ for a fixed $Ra$. 

Regular patterns in the turbulent convection regime were studied in detail by Fitzjarrald (1976) in a square 
cell filled with air for aspect ratios between 2 and 58 covering a range of Rayleigh numbers between 
$4\times 10^4$ and $7\times 10^9$.  He calculated the dominant horizontal scales from the Fourier co-spectra 
of $u_z$ and $T$. The spectral peak in the heat flux corresponds to a wavelength $\Lambda$ that increased 
from $4 H$ to $6 H$ for $4\times 10^4<Ra<1.7\times 10^7$ and thus  $58>\Gamma>15$. Based on 
Figs. \ref{fig5} and \ref{fig6}  for $Ra=10^7$ and Fig. \ref{fig7} for $Ra=10^8$, we take the width of 
the large-scale circulation roll (which corresponds to the spacing 
between local maxima of $\langle u_z\theta\rangle$) as $\Lambda/2$ and get thus a wavelength 
$\Lambda\approx 4 H$ for $\Gamma=6, 8$ and 12. The associated wavenumber $k=2\pi/\Lambda\approx 1.5 H^{-1}$
which is about half the size of $k_c=3.117 H^{-1}$ at the onset of thermal convection in an 
infinite layer. The dominant horizontal scales are similar to those of Fitzjarrald. The results in Fig.~5 further confirm the observation made by Fitzjarrald. This wavelength shrinks at smaller aspect ratios where the pattern 
has to fit into the cylindrical cell. Hartlep {\it et al.} (2005) have also traced back their large-scale turbulent 
temperature patterns to the states which are observed in the weakly nonlinear regime. A series of simulations at 
$\Gamma=10$ for Rayleigh numbers up to $Ra=10^7$ confirms a characteristic wavelength of half their box size, i.e. $\Lambda\approx 
5H$ for $Pr=0.7$. This wavelength was $\Lambda\approx 3H$ at $Ra=4000$. Their study 
shows in addition a clear  shape dependence of the circulation rolls on the Prandtl  number. Slight variations of 
$\Lambda$ in the three studies might be caused by different cell geometries and boundary conditions in the 
simulations. Nevertheless, the same range of wavelengths can be observed for $Ra\sim 10^7$ and $Pr=0.7$ in 
all works. 

Although qualitative similarities between the LSC patterns at small $Ra$ and those at higher $Ra$ 
are obvious from Fig.~\ref{fig6}, we can expect that the particular mechanisms that drive the large-scale flow will be different. 
The onset of a flow motion for small $Ra$  is triggered by a slight  dominance of buoyancy forces per 
unit mass, $f_b=g\alpha \theta$, compared to the restoring drag forces per unit mass, $f_d=\frac{1}{2}
C_f u_z^2 H$. This is the simple chaotic waterwheel picture by Malkus and Howard (see Strogatz 1994).   
In the turbulent case, the heat transport through the thin thermal boundary layers is responsible for 
large-scale spatial temperature differences.  Spatial temperature differences create pressure gradients 
which drive the large-scale flow (Reeuwijk {\it et al.} 2008).   This might the reason  why the wavelength of the circulation 
rolls is slightly increasing with growing $Ra$. 

We can summarise  that, for the range of parameters covered here, the LSC patterns do not disappear in the turbulent 
regime up to $Ra=10^9$. For the larger aspect ratios pentagon-like circulation cells are formed preferentially.

\begin{figure}
\begin{center}
\includegraphics[width=6.5cm]{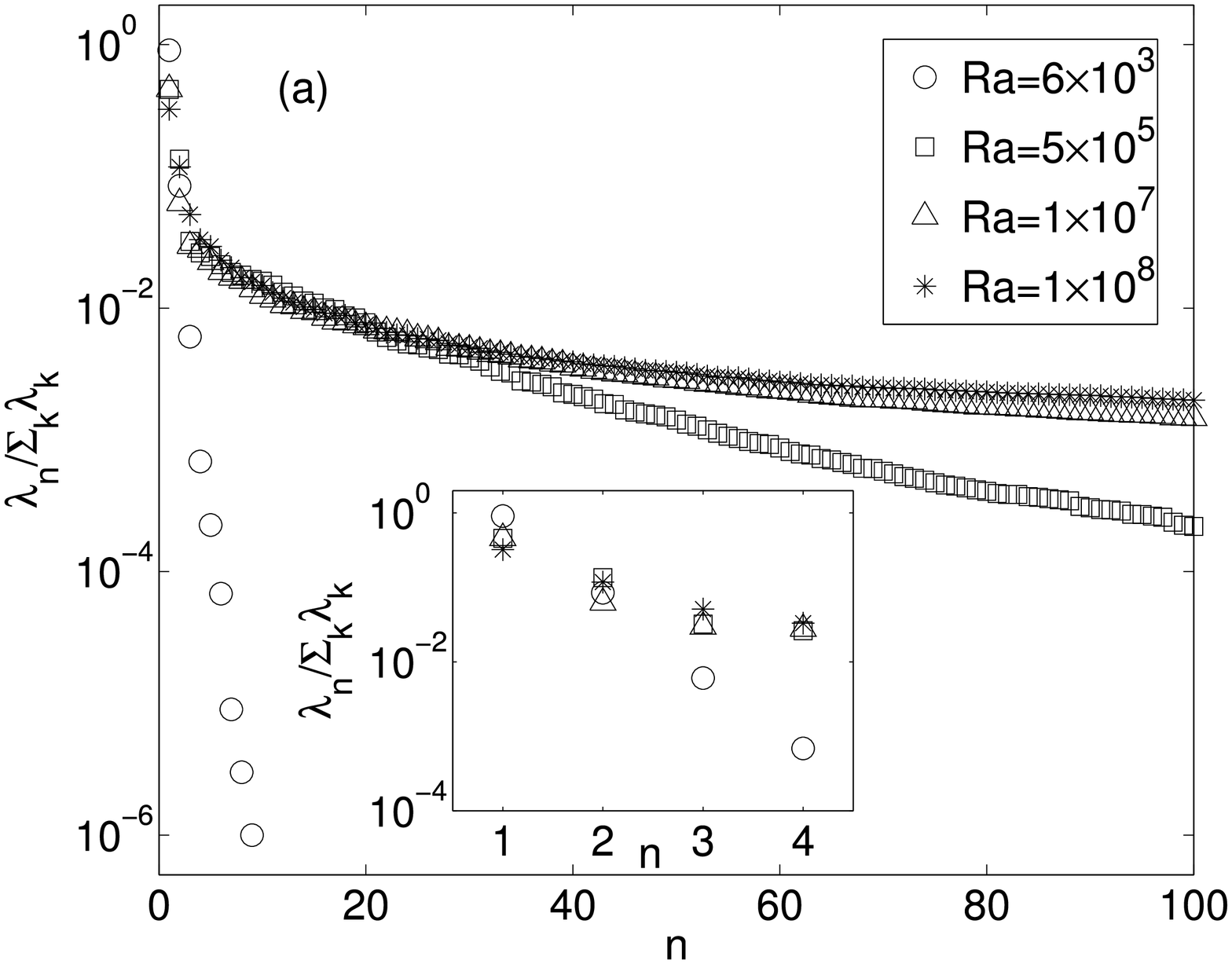}
\includegraphics[width=6.5cm]{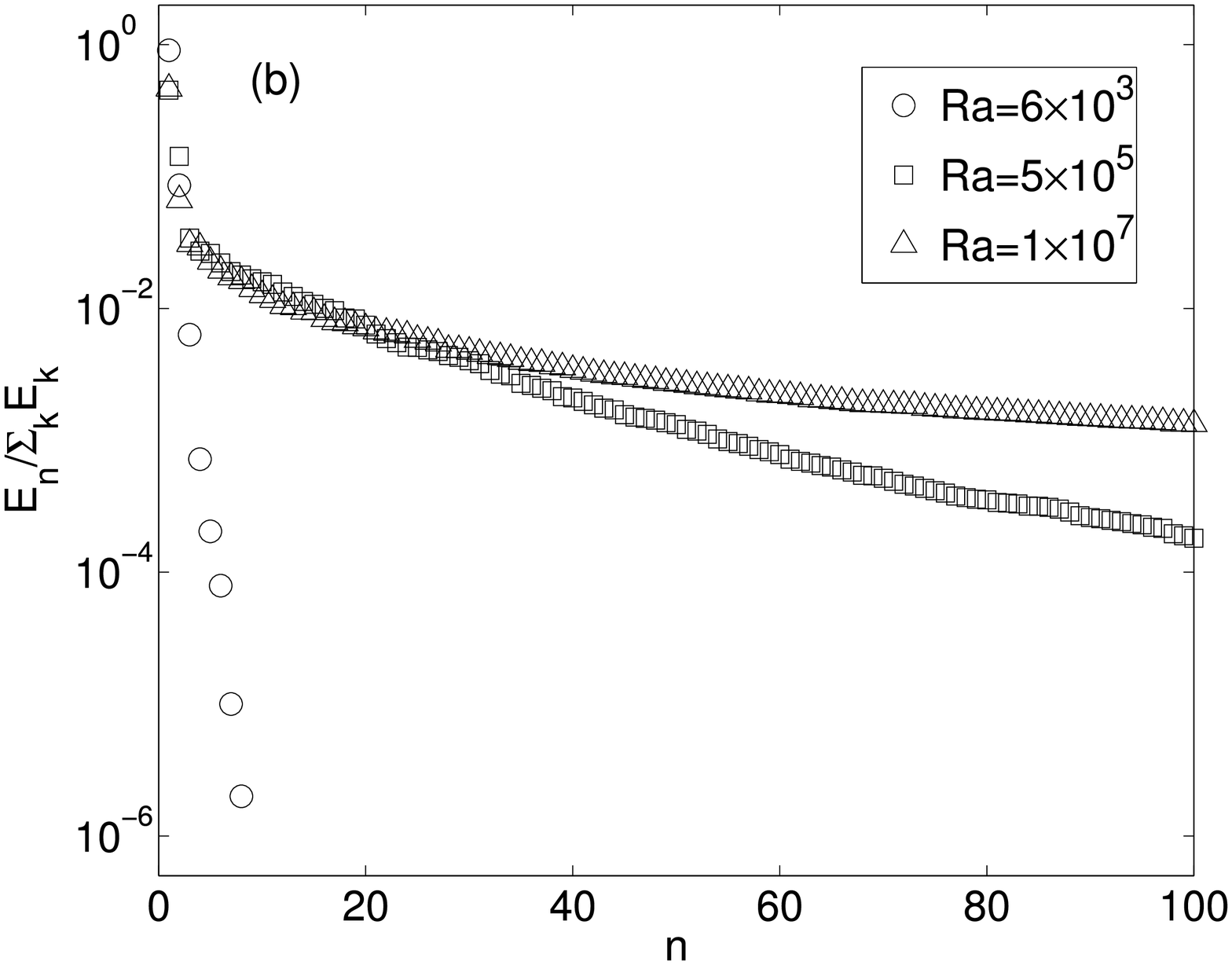}
\caption{(a) Normalized POD eigenvalue spectrum of the total energy (thermal plus kinetic) is shown for four different Rayleigh numbers as indicated in the legend. (b) Normalized POD eigenvalue spectrum of the kinetic energy is shown for three different Rayleigh numbers as indicated in the legend. The snapshot method was conducted over 100 state vectors. The aspect ratio is $\Gamma=3$ in all four cases.
The spectra in (a) and (b) coincide almost perfectly. The inset magnifies the spectra for the first few modes.} 
\label{fig8}
\end{center}
\end{figure}
\section{Proper Orthogonal Decomposition of the turbulent convection flow}
\subsection{The snapshot method}
The turbulent heat transfer is the sum of transfers by the LSC and the turbulent fluctuations. In order 
to disentangle both contributions systematically, we conduct a so-called Karhunen-Lo\`{e}ve method or Proper 
Orthogonal Decomposition (POD). The reader is referred to Smith {\it et al.} (2005) for a compact tutorial 
on this subject. Here, we outline the basic ideas only. The application of the POD method to the convection 
problem goes primarily back to Sirovich and his co-workers (see e.g.  Sirovich \& Park 1990). Consider a 
state vector ${\bm v}=({\bm u},\theta)$ with zero mean, $\langle {\bm v}\rangle=0$. It has a mean turbulent 
energy (kinetic energy plus temperature variance), which is given by
\begin{equation}
E=\langle ({\bm v},{\bm v})\rangle= \Big\langle \int_V \sum_{k=1}^4 v_k({\bm x},t) v_k({\bm x},t)\,\mbox{d}^3x 
\Big\rangle_t\,,
\end{equation}
where $V$ is the cell volume and the scalar product $(\cdot,\cdot)$ is defined in $L_2(V$). At the core 
of the method is the determination of the POD modes ${\bm \phi}({\bm x})$, which maximize the following 
functional
\begin{equation}
\frac{\langle|({\bm v},{\bm \phi})|^2\rangle_t}{({\bm \phi},{\bm \phi})}\rightarrow \mbox{max}.
\end{equation}
Variational calculus then yields the following integral equation 
\begin{equation}
\int_V \hat{\bm K}({\bm x}, {\bm x}^{\prime}) {\bm \phi}^{(m)}({\bm x}^{\prime})\,\mbox{d}^3x^{\prime}=
\lambda_m {\bm\phi}^{(m)}({\bm x}),
\label{integraleq}
\end{equation}
with the kernel (or covariance matrix) $K_{ij}({\bm x}, {\bm x}^{\prime})=\langle v_i({\bm x},t) v_j
({\bm x}^{\prime},t)\rangle_t$ and $i,j=1,2,3,4$. If the kernel is a Hermitian and non-negative operator, the set of 
empirical eigenfunctions $\{ {\bm \phi}^{(m)}\}$ forms an orthonormal system, i.e. 
$({\bm \phi}^{(m)}, {\bm \phi}^{(n)})=\delta_{mn}$. The integral equation is transformed into a
matrix eigenvalue problem. In our case the size of the kernel becomes extremely large, namely a 
$4N\times 4N$ matrix for ${\bm v}=({\bm u}, \theta)$ and $N=N_r\times N_{\phi}\times N_z$. 
Symmetries and incompressibility of the flow reduce the number of degrees of freedom in many 
cases. However, we still have to apply the method of snapshots, which is the preferred 
choice if $N\gg N_T$, with $N_T$ the number of snapshots. We therefore construct empirical 
eigenfunctions as a linear combination of the state vectors ${\bm v}$, where the eigenfunctions are 
given by
\begin{equation}
{\bm \phi}^{(m)}({\bm x})=\sum_{i=1}^{N_T} \alpha_i^{(m)} {\bm v}({\bm x},t_i)\,.
\label{integraleq1}
\end{equation}
Such a procedure reduces the complexity of the problem and leads to the solution of 
an eigenvalue problem of $N_T\times N_T$ matrix, as is evident from the subsequent 
expressions. If $\langle\cdot\rangle_t$ is substituted by an arithmetic  mean over the snapshots, 
it follows from (\ref{integraleq}) that 
\begin{equation}
\frac{1}{N_T}\int_V \sum_{k=1}^{N_T}v_p({\bm x},t_k)v_n({\bm x}^{\prime},t_k) 
\phi_n^{(m)}({\bm x}^{\prime})\,\mbox{d}^3x^{\prime}=
\lambda_m \phi_p^{(m)}({\bm x}).
\label{integraleq2}
\end{equation}
With (\ref{integraleq1}) one can arrive at
\begin{equation}
\sum_{k=1}^{N_T}v_p({\bm x},t_k) \left[ \sum_{i=1}^{N_T} \frac{1}{N_T} \int_V v_n
({\bm x}^{\prime},t_k) v_n({\bm x}^{\prime},t_i)\,
\mbox{d}^3x^{\prime}\,\alpha_i^{(m)} \right]=\sum_{q=1}^{N_T} v_p({\bm x},t_q) 
\lambda_m \alpha_q^{(m)}\,,
\label{integraleq2a}
\end{equation}
and thus 
\begin{equation}
\sum_{i=1}^{N_T} \frac{1}{N_T} ({\bm v}(t_k), {\bm v}(t_i))\alpha_i^{(m)} =
\sum_{i=1}^{N_T} C_{ki}\alpha_i^{(m)}= 
\lambda_m \alpha_k^{(m)}\,.
\label{integraleq3}
\end{equation}
Eventually, $N_T$ eigenvectors $\{{\bm\alpha}^{(m)}\}$, with $m=1,2,...,N_T$, represent
$N_T$ POD modes $\{{\bm\phi}^{(m)}\}$ (vectors of $4N$ components) constructed 
from the state vectors. 

We proceed in two different steps. First, we use ${\bm v}={\bm u}$ only and not the combined 
velocity-temperature state  vectors. The eigenvalue spectrum $E_1\ge E_2
\ge \dots E_{N_T}$ then quantifies  the fraction of the turbulent kinetic energy contained 
in each of the $N_T$ POD modes. Second, we use ${\bm v}=({\bm u},\theta)$ and determine 
the total energy spectrum $\lambda_1\ge \lambda_2\ge \dots \lambda_{N_T}$. The latter will 
be used in the subsequent sections. Both eigenvalue spectra are presented in Fig.~\ref{fig8} 
for different Rayleigh numbers and $N_T=100$ snapshots. 
At the smallest Rayleigh number $Ra=6\times 10^3$, the first few POD modes contain most of 
the total energy (Fig. \ref{fig8}(a)) and kinetic energy (Fig. \ref{fig8}(b)).  This is the weakly nonlinear 
regime of convection. With increasing Rayleigh number, the spectra decay slowly. 
For Rayleigh numbers $Ra\ge 10^7$ the convection is turbulent and  a significant fraction of 
the kinetic and total energy is distributed among the higher-order POD modes.  The inset in Fig.~\ref{fig8}(a) shows the magnified view for the first few POD modes. 
This observation is in agreement with Sirovich \& Park (1990). The dynamic significance of the 
subsequent modes increases steadily with increasing Rayleigh number since turbulent fluctuations 
are present. 

\subsection{Spatial structure of primary and secondary modes}
\begin{figure}
\begin{center}
\includegraphics[width=6.5cm]{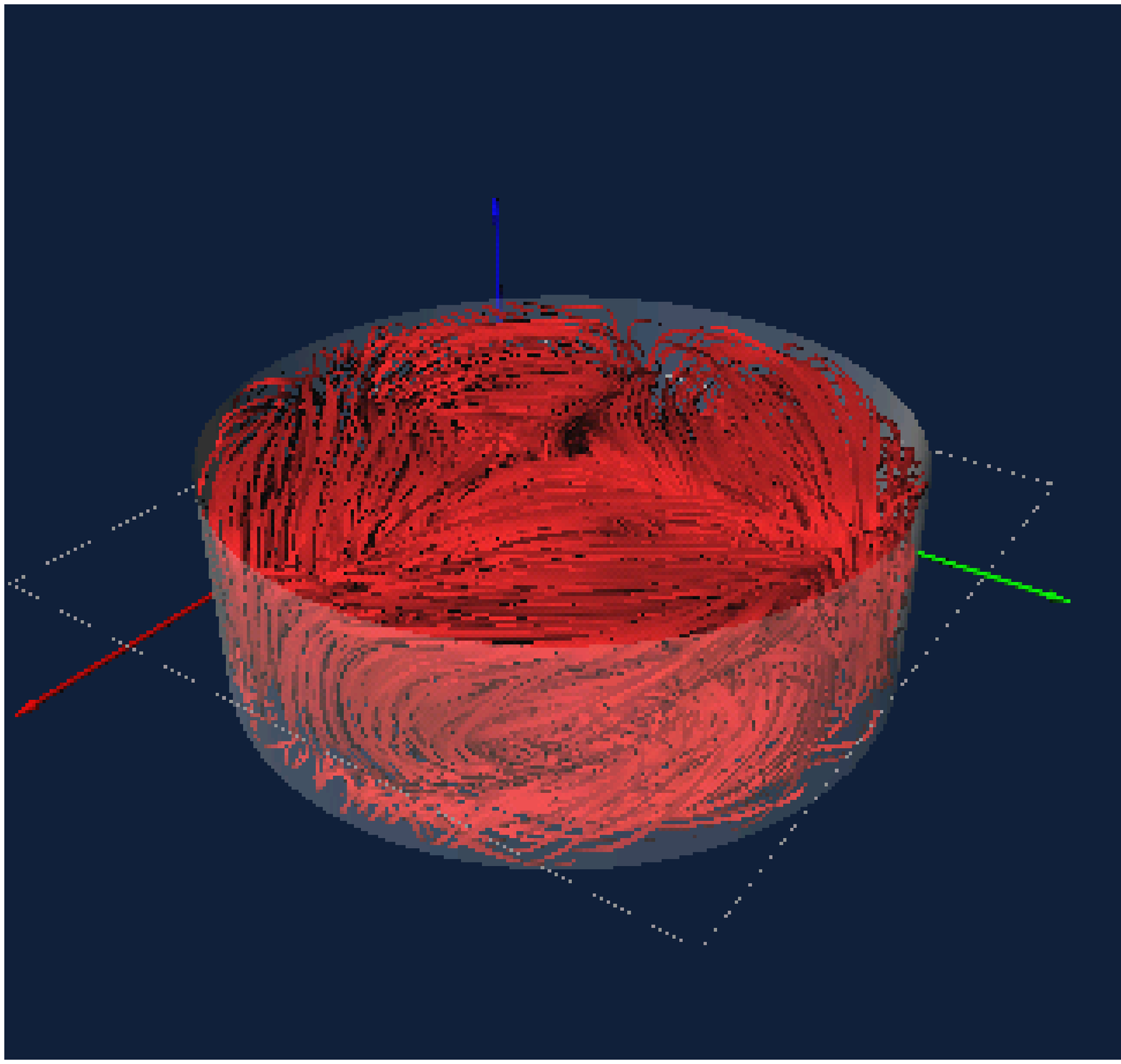}
\hspace{0.2cm}
\includegraphics[width=6.5cm]{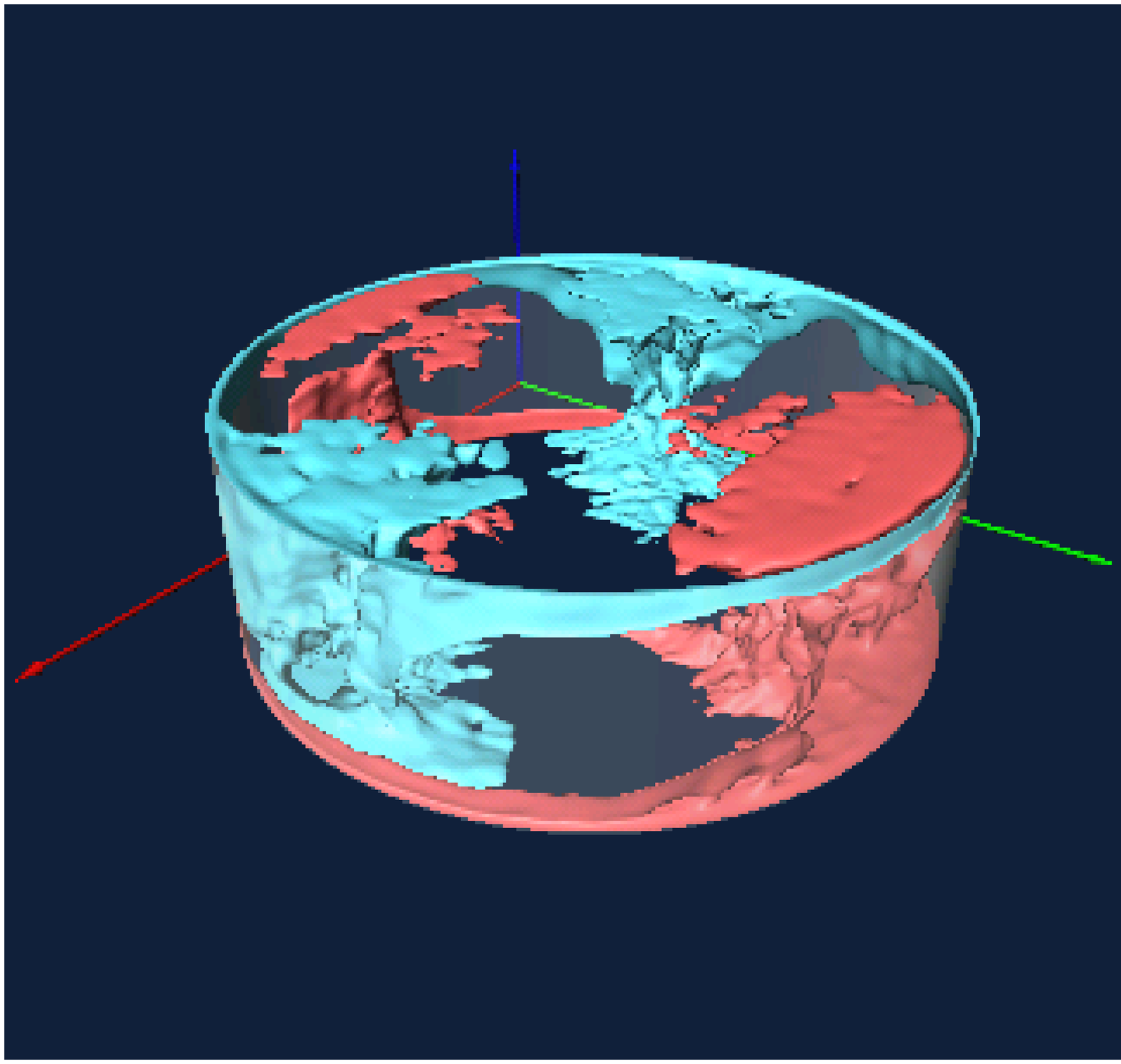}

\vspace{0.2cm}
\includegraphics[width=6.5cm]{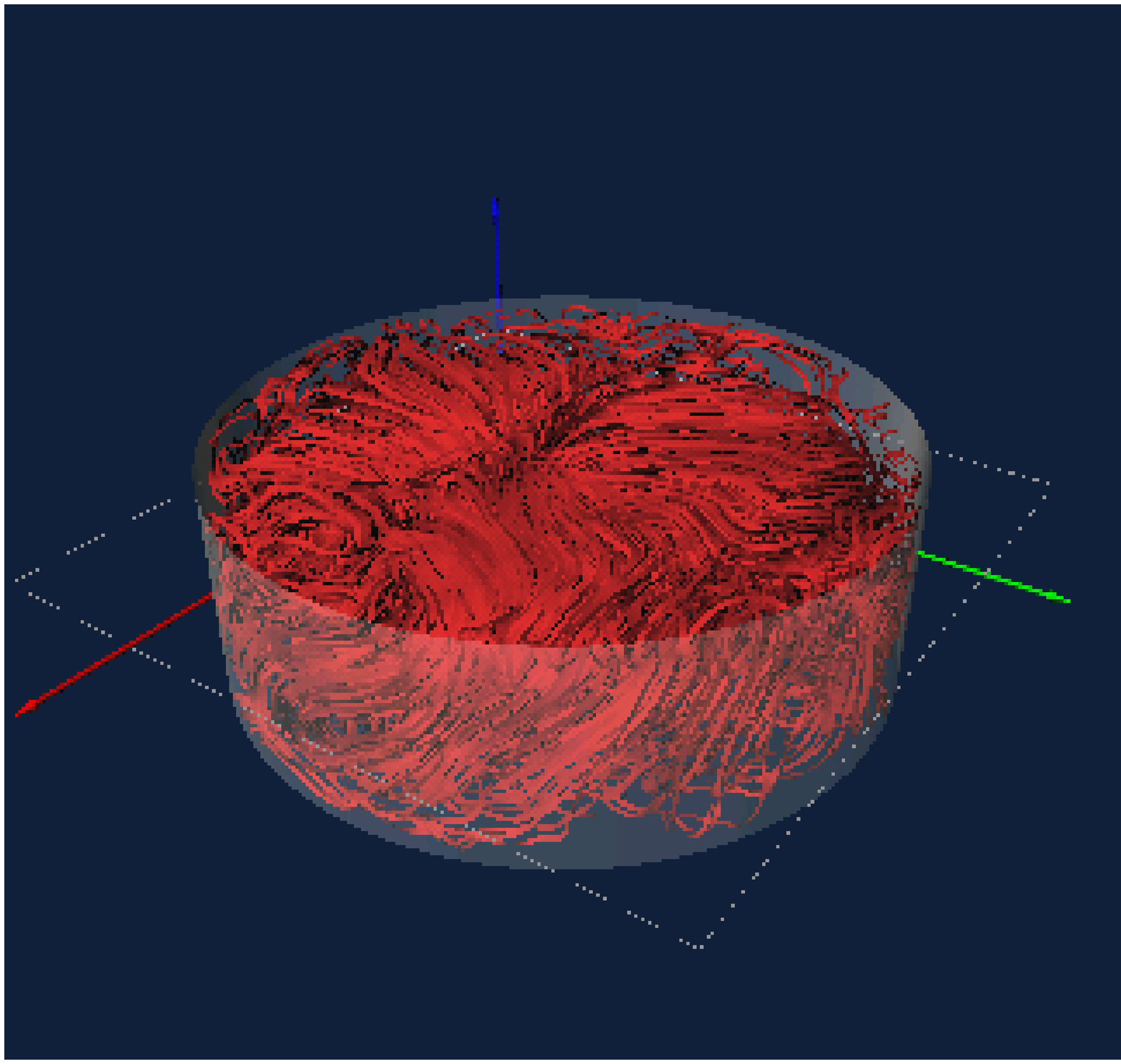}
\hspace{0.2cm}
\includegraphics[width=6.5cm]{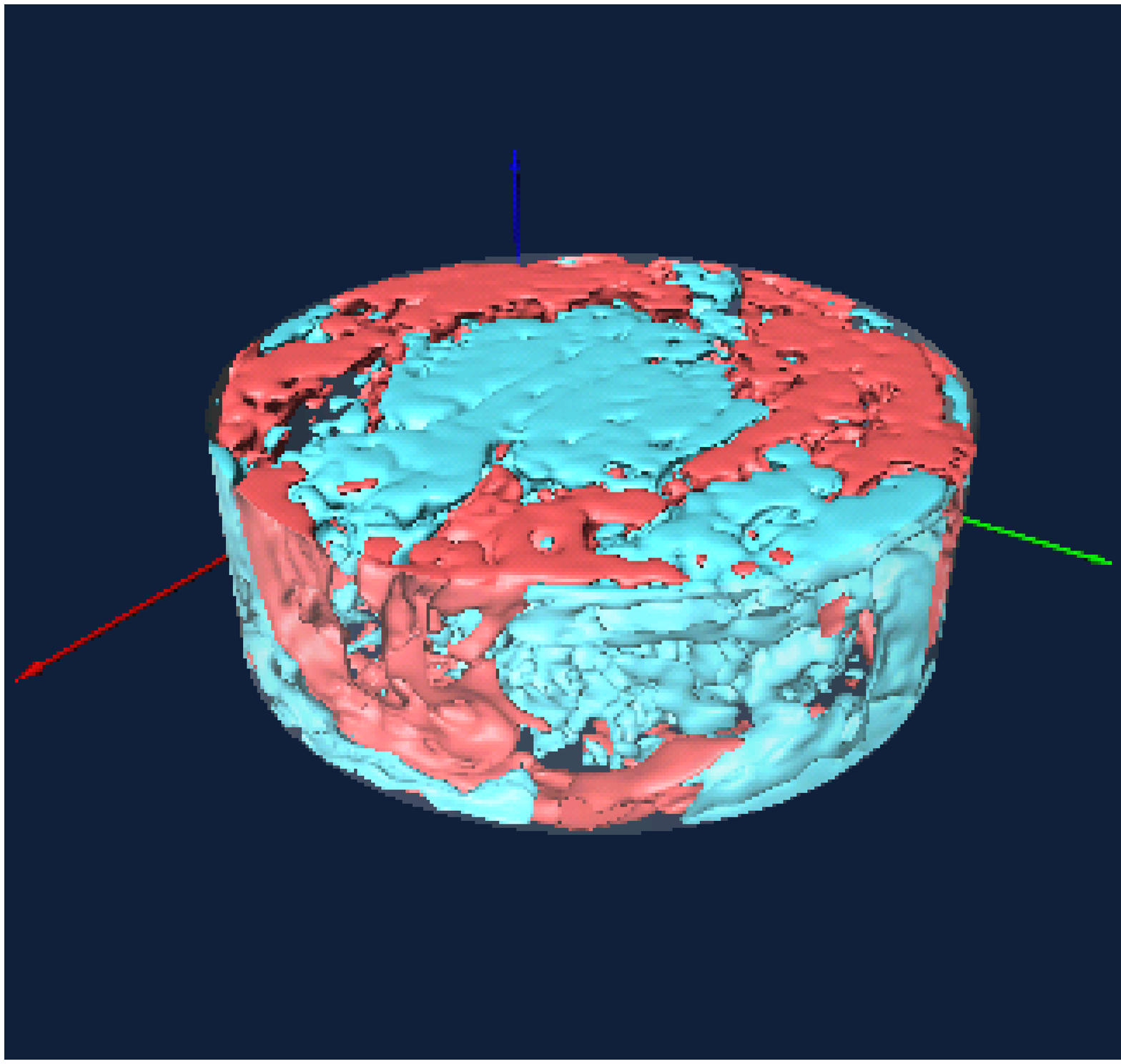}
\caption{Three-dimensional visualization of the first two POD modes for $Ra=10^7$ and 
$\Gamma=3$. Upper row:  Streamlines of the primary flow mode $(\phi_1^{(1)}, \phi_2^{(1)}, 
\phi_3^{(1)})$ (left) and isosurfaces of the primary temperature mode $\phi_4^{(1)}$ (right) 
at the isolevels $\pm 0.5\Delta T$. Lower row:  Streamlines of the secondary flow mode 
$(\phi_1^{(2)}, \phi_2^{(2)}, \phi_3^{(2)})$ (left) and isosurfaces of the secondary temperature 
mode $\phi_4^{(2)}$ (right) at the isolevels  $\pm 0.095\Delta T$. Blue isosurfaces correspond 
to negative values and red isosurfaces to positive values in both figures.} 
\label{fig9}
\end{center}
\end{figure}

Before we proceed to the analysis of the turbulent heat transfer, we visualize the spatial structure 
of the first two POD modes and compare it with the LSC. Fig.~\ref{fig9} shows the three-dimensional 
view of the primary and secondary modes. The velocity field ($\phi_1^{(m)}({\bm x}), \phi_2^{(m)}({\bm x}), 
\phi_3^{(m)}({\bm x}))$ with $m=1, 2$, is plotted as streamlines in the left column and the temperature 
field $\phi_4^{(m)}({\bm x})$ at two isolevels is plotted in the right column. The data set corresponds to 
$\Gamma=3$ and $Ra=10^7$. The structure of the velocity field of the primary mode almost exactly 
replicates the time-averaged velocity field shown in Fig.~\ref{fig5}. This replication is also verified for 
other aspect ratios, which are not shown here. The shape of the primary temperature POD mode indicates 
hot up- and cold  downwellings on the side wall. The two lower panels of Fig. \ref{fig9} show that the
secondary modes exhibit a more complex structure. The primary and secondary POD mode have the
same number of large-scale rolls. In addition, we detect  smaller substructures of the secondary modes, such 
as recirculation vortices close to the top and bottom plates and weak modulations of the large-scale rolls.     

\subsection{Heat transfer by different POD modes}
The contribution of different subsets of the POD modes to the turbulent heat transfer is determined as 
follows. We can decompose turbulent snapshots as
\begin{eqnarray}
u_i({\bm x},t)&=&\sum_{m=1}^{N_T} a_m(t) \phi_i^{(m)}({\bm x})\,,\\
\theta({\bm x},t)&=&\sum_{m=1}^{N_T} a_m(t) \phi_4^{(m)}({\bm x})\,,
\label{POD1}
\end{eqnarray}
with $i=1, 2, 3$ (or $x, y, z$). The coefficients $a_m(t)$ correspond to the projection of the turbulent 
flow field at time $t$ to mode ${\bm \phi}^{(m)}({\bm x})$, which are calculated from the scalar product 
in $L_2(V)$.  The Nusselt number definition (\ref{Nug}) then translates to
\begin{eqnarray} \nonumber
Nu(N_T)&=&1+\frac{H}{\kappa\Delta T} \sum_{m,n=1}^{N_T} \Big\langle a_m(t) \phi_3^{(m)}({\bm x}) 
\left[ \overline{T}(z)+ a_n(t) \phi_4^{(n)}({\bm x})\right]\Big\rangle_{V,t}\\ \nonumber
&=&1+\frac{H}{\kappa\Delta T} \sum_{m,n=1}^{N_T} \Big\langle a_m(t) \phi_3^{(m)}({\bm x}) 
a_n(t) \phi_4^{(n)}({\bm x})\Big\rangle_{V,t}\\ 
&=&1+\frac{H}{\kappa\Delta T} \sum_{m=1}^{N_T}\lambda_m\Big\langle \phi_3^{(m)}({\bm x}) \phi_4^{(m)}({\bm x})\Big\rangle_{V}\,, 
\label{POD2}
\end{eqnarray}
where $\overline{T}(z)=\langle T(z)\rangle_{A,t}$. The contribution of the mean profile drops out. 

\begin{figure}
\begin{center}
\includegraphics[width=6.5cm]{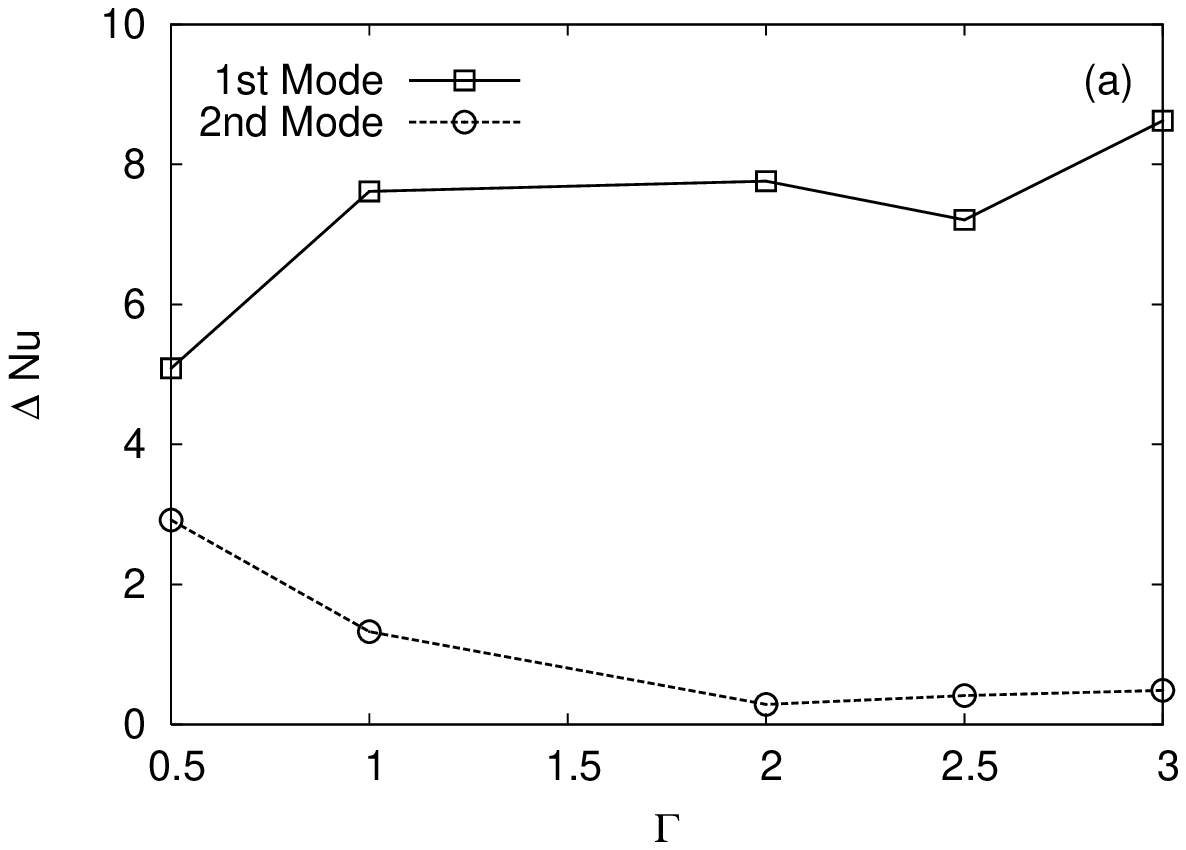}
\includegraphics[width=6.5cm]{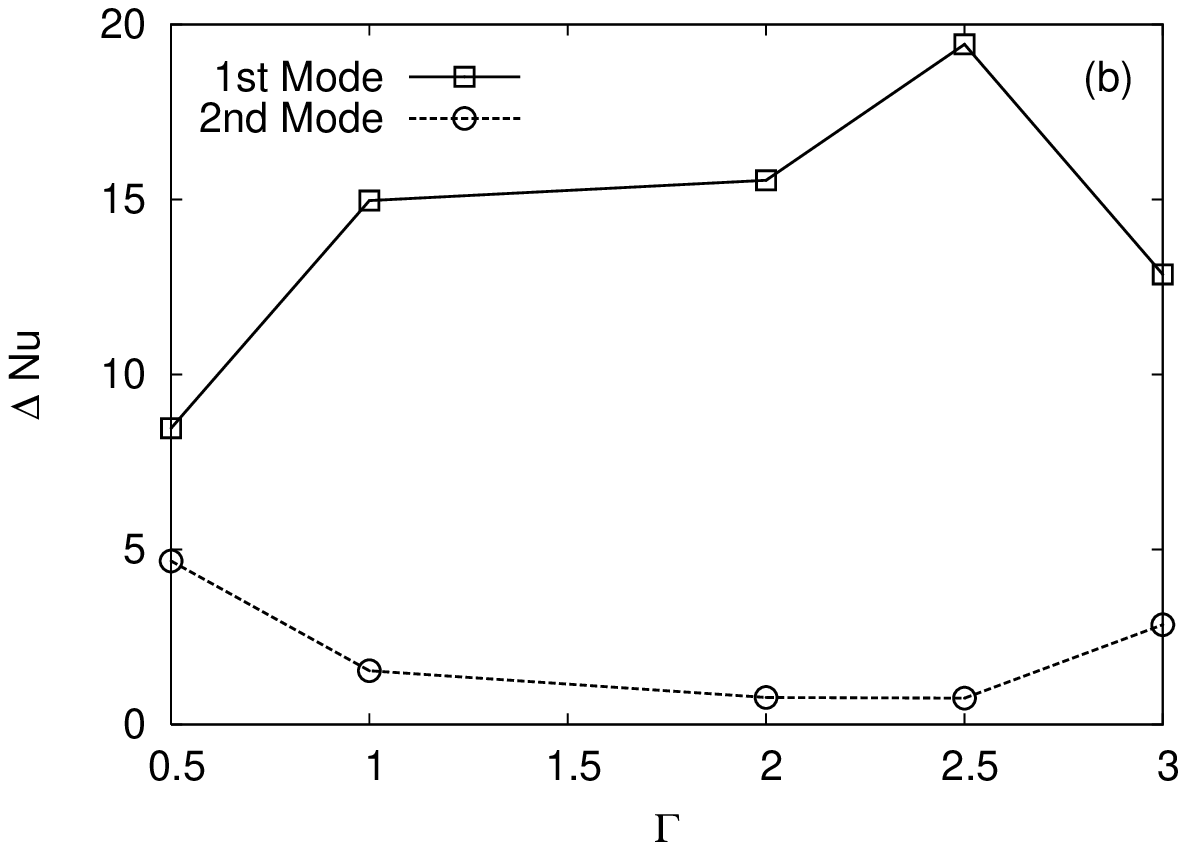}
\vspace{0.2cm}
\includegraphics[width=6.5cm]{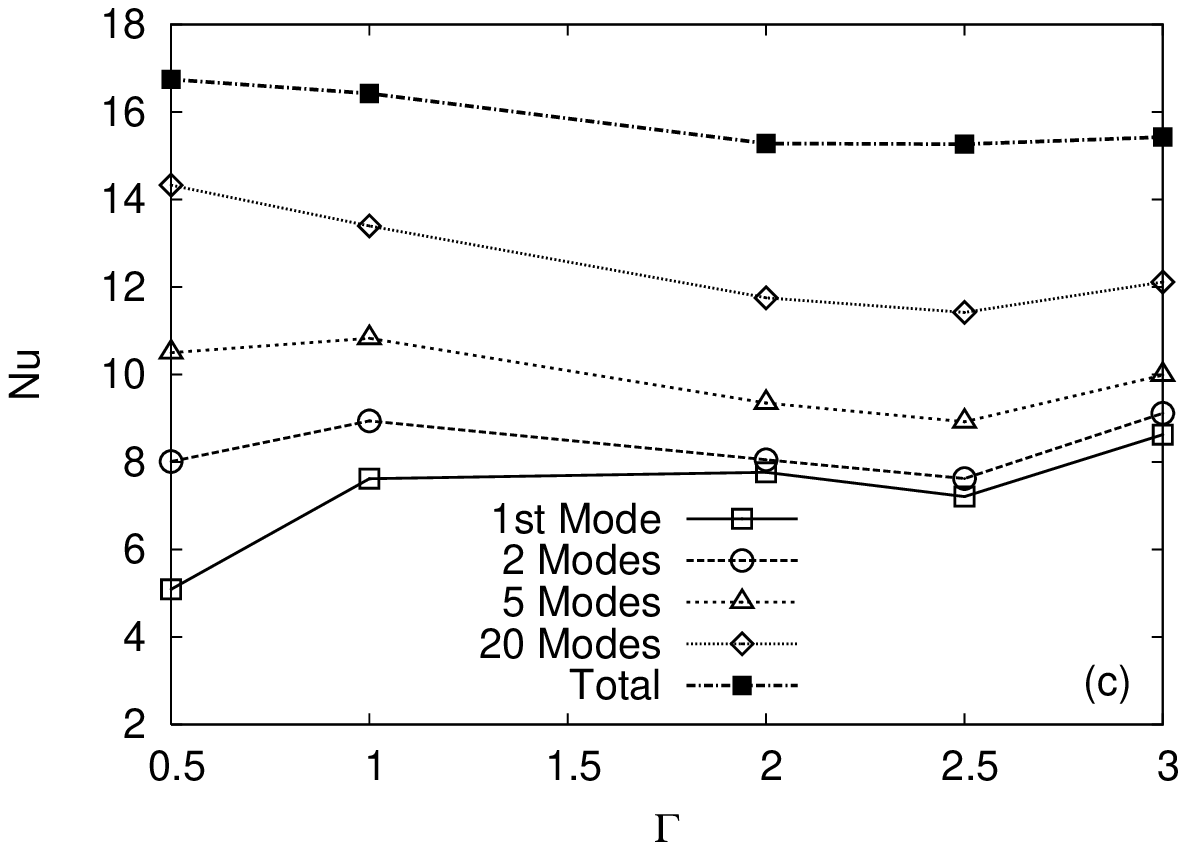}
\includegraphics[width=6.5cm]{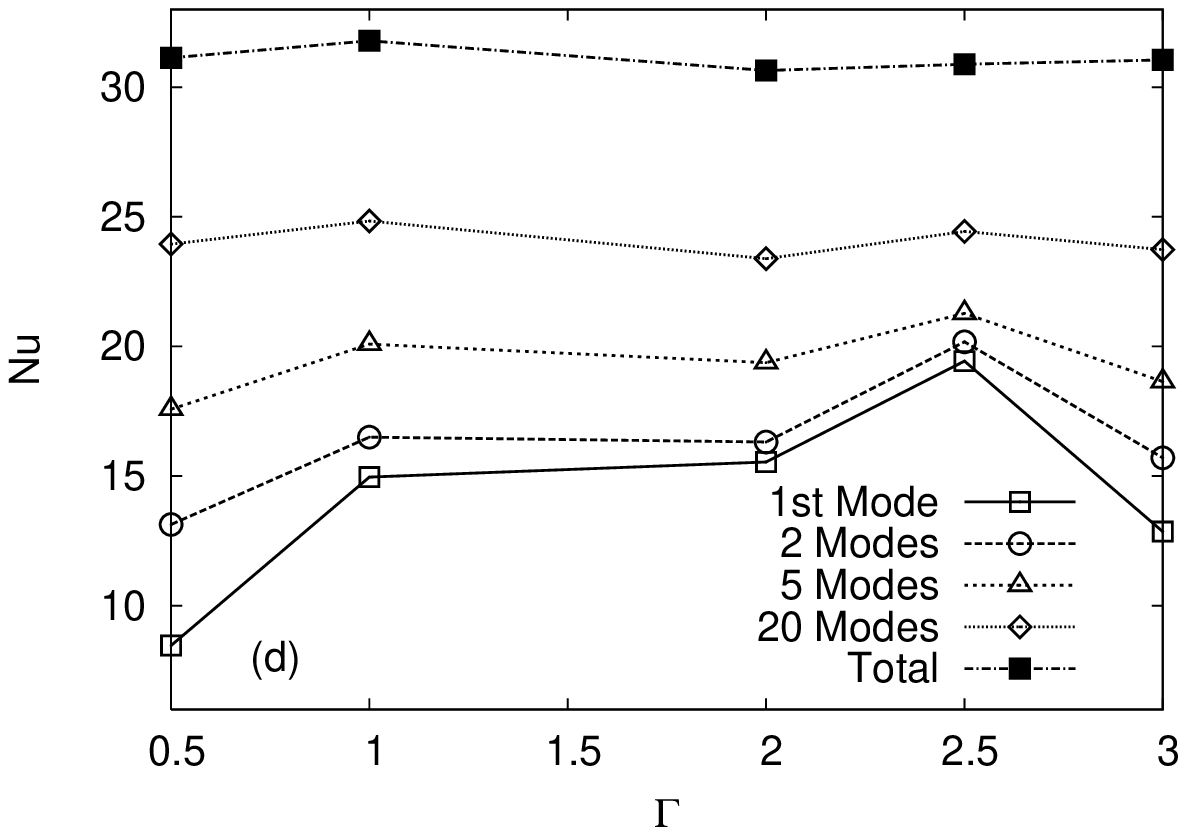}
\caption{Contribution of various POD modes (as indicated in the legend) to the global heat 
transfer for two Rayleigh numbers and five different aspect ratios. (a) Contribution of the 
primary and secondary modes for $Ra=10^7$. (b) Contribution of the 
primary and secondary modes for $Ra=10^8$. (c) Accumulated contributions for $Ra=10^7$.
(d) Accumulated contributions for $Ra=10^8$. For completeness we also add the original Eulerian 
values of the Nusselt number.} 
\label{fig10}
\end{center}
\end{figure}

In Fig.~\ref{fig10}, we report the contribution of 
various POD modes to the global heat transfer for $Ra=10^7$ and $10^8$. The contribution of the primary 
and secondary modes is displayed in panels (a) and (b). The expansion (\ref{POD2}) is then truncated after 
1, 2, 5, 20, and 100 POD modes.  Panels (c) and (d) show the accumulated fraction to the heat transfer for the
number of modes as given in the legend of both figures. 
As a consistency check, we compare the full expansion 
which is based on 100 snapshots with the Eulerian value as determined in section 3. 
Computational resources limit the present analysis to $N_T=100$ 
since intensive data in- and output is required. The values of $Nu$ still deviate slightly from those in Tab. 1. 
It is found that the convergence is slow in particular for the higher-order modes. The slow convergence was also 
underlined in Fig.~\ref{fig3}(d).  In Tab. 3 we have listed in addition
some quantitative details of the POD analysis of the heat transfer. 
The results are consistent with the Eulerian values in Tab. 1.

The primary POD mode carries  the following fraction of the global heat transfer
\begin{equation}
Nu(N_T=1)=1+\frac{\lambda_1 H}{\kappa\Delta T} \Big\langle \phi_3^{(1)} \phi_4^{(1)}\Big\rangle_{V}\,.
\end{equation}
\begin{table}
\begin{center}
\begin{tabular}{ccccccccccc}
\multicolumn{1}{c}{$Ra$}       & \multicolumn{5}{c}{$10^7$} &\multicolumn{5}{c}{$10^8$} \\
$\Gamma$                               & 0.5     & 1.0      & 2.0      & 2.5      & 3.0       & 0.5      & 1.0     & 2.0      & 2.5      & 3.0   \\
$Nu$                                         & 17.08 & 16.73 & 15.88 &  15.77 & 16.06  & 32.06 & 32.21 & 31.25 & 31.87 & 32.29   \\
$Nu(N_T=100)$                     & 16.74 & 16.42 & 15.28 &  15.26 & 15.42  & 31.13 & 31.78 & 30.64 & 30.88 & 31.05   \\
$\frac{Nu-Nu(N_T=100)}{Nu}$  & 2.0\% & 1.8\% & 3.8\% &  3.3\% & 4.0\%  & 2.9\% & 1.3\% & 1.9\% & 3.1\% & 3.8\%   \\
$\frac{Nu(N_T=1)}{Nu(N_T=100)}$  & 30\% & 46\% & 51\% &  47\% & 55\%  & 27\% & 47\% & 51\% & 63\% &  41\%  \\
\end{tabular}
\label{tab3}
\caption{Turbulent heat transport of POD modes. The Nusselt number $Nu$ of the analysis of the DNS 
data (taken from Table 1) is compared with that obtained from a sequence of $N_T=100$ snapshots 
denoted by $Nu(N_T=100)$. Furthermore, the contribution of the primary mode, $Nu(N_T=1)$, 
is compared to the total transport, $Nu(N_T=100)$.}
\end{center}
\end{table}
\begin{figure}
\begin{center}
\includegraphics[width=6.5cm]{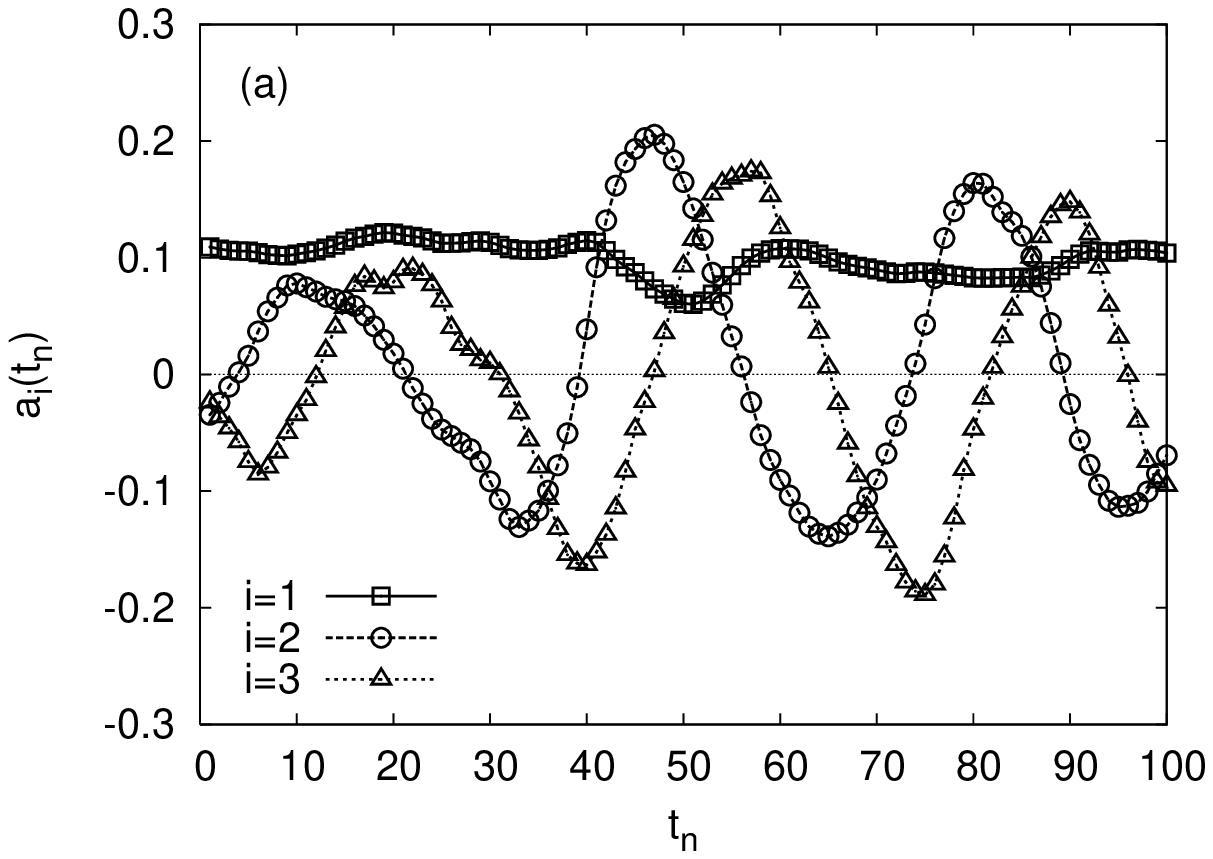}
\hspace{0.2cm}
\includegraphics[width=6.5cm]{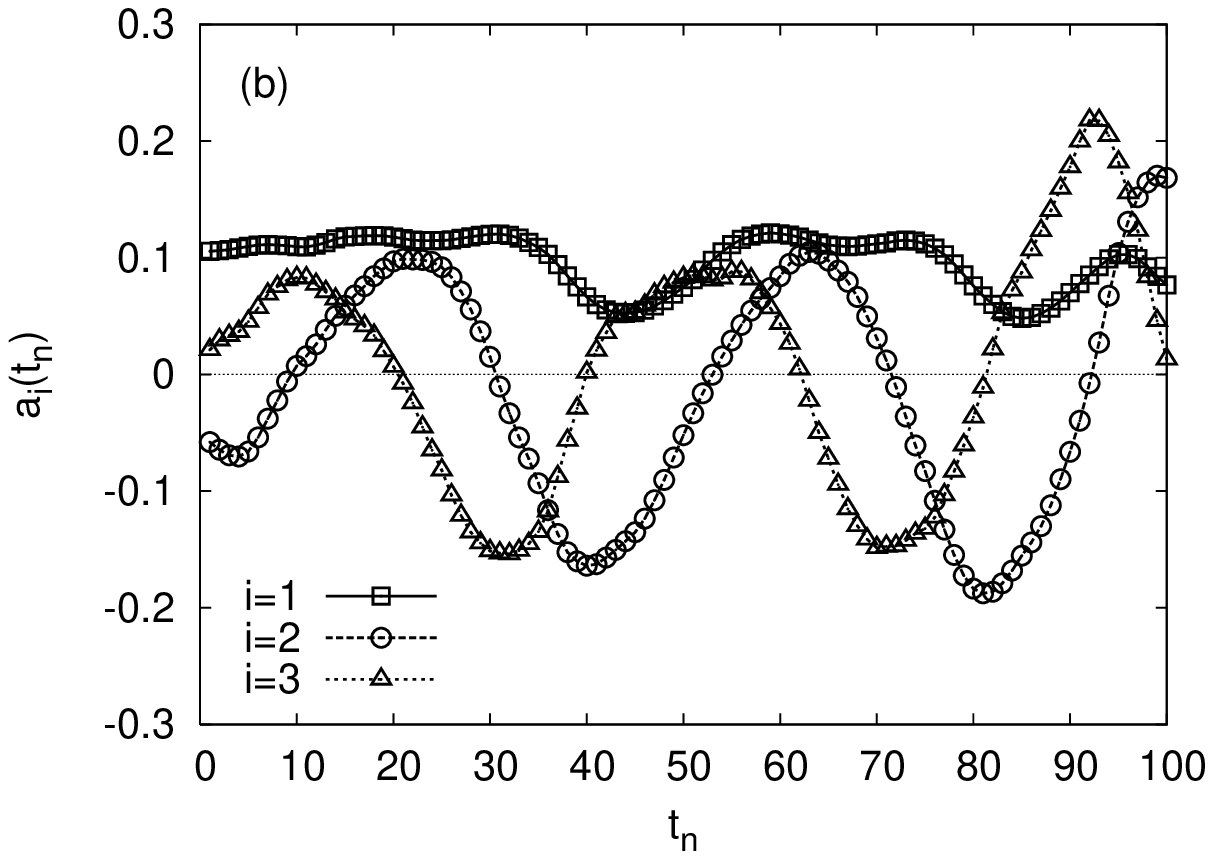}
\caption{Time dependence of $a_i(t)$ for $\Gamma=2$. The snapshots are therefore projected on the 
primary $(i=1)$, secondary $(i=2)$, and tertiary mode $(i=3)$.  (a) $Ra=10^7$. (b) $Ra=10^8$.} 
\label{fig11}
\end{center}
\end{figure}
For flow patterns with a single-roll circulation,  i.e. $\Gamma=$ 1, 2, and 2.5 for $Ra=10^7$ and 
$\Gamma=$1 and 2 for $Ra=10^8$, the contribution to the heat transfer by the primary POD mode 
is about the same. It makes up about one half of the total amount. This contribution increases by 
10\% due to the transition from a single-roll to a double-roll pattern between $\Gamma=$ 2.5 and 3 
for $Ra=10^7$. The double-roll LSC can carry more heat through the cell since the number of 
up- and downwelling regions with $\langle u_z\theta\rangle_t >0$ increases across the cell. This 
can also be seen in the plots in the lower row of Fig.~\ref{fig5}. One can consider the dynamics 
around $\Gamma=2.5$ as a bottleneck for the heat transfer. The one-roll pattern gets 
ever flatter with increasing $\Gamma$ and can thus transfer heat less efficiently through the cell. 
Once the two-roll pattern is established, this bottleneck is removed and the share of the primary 
mode in the total heat transfer increases. The same transition appears between aspect ratios of 2 
and 2.5 for $Ra=10^8$. Again, we detect a jump of the primary mode contribution by 12\%. The 
opposite is the case for the slender cell at $\Gamma=0.5$. One observes a much lower $Nu$ 
fraction due to the primary mode in comparison to the cases with $\Gamma\ge1$. This can be 
attributed to the complex flow configuration  in the slender cell, in which there are either two 
counter-rotating rolls on top of each other, or one slender roll  (Verzicco \& Camussi 2003, Xi \& 
Xia 2008a).

The secondary and higher-order modes provide information that can be obtained within the present 
POD analysis only. The fraction of the secondary POD mode (see Fig.~\ref{fig10} (a) and (b)) to the 
global heat transfer is much smaller than that of the primary. It is about 5\% for the larger aspect 
ratios and remains almost insensitive when the primary mode switches from a one-roll to a two-roll 
pattern. A closer inspection of both plots suggests, however, that an increase of the portion of the total heat transfer due to the primary mode causes a decrease of that of the secondary mode. This is 
clearly indicated for $\Gamma=0.5$ and 1 in both $Ra=10^7$ and $10^8$ series, and for $\Gamma=2$, 2.5 and 3 in the 
series with $Ra=10^8$. It further supports our arguments in the last paragraph. The local minimum of the secondary POD mode contribution coincides with the local maximum of the primary one. When the primary mode becomes less efficient in transferring heat, the secondary mode has to take a bigger share. 
The two panels in Fig.~\ref{fig11} display finally the time dependence of the expansion coefficients 
of the first three POD modes, $a_m(t)$ with $m=1,2,3$. The graphs are obtained by projecting the 100 snapshots
onto the POD modes $\phi_i^{(m)}({\bm x})$ for $m=1,2,3$. While the primary modes remains nearly 
constant, we see that the secondary and tertiary modes oscillate with a period of  approximately 
$30 t_f$ and are shifted with respect to each other by about $10 t_f$.  The secondary and tertiary  
mode contribute thus mainly to temporal variance of the heat transfer. Their time-averaged contributions 
remain, however, significantly lower than that of the primary mode.
\section{Summary and discussion}   
Within the parameter range of the present study, the DNS results have revealed a dependence of the 
Nusselt number on the aspect ratio. The variation in $Nu(\Gamma)$ curve is between 11\% and 3\%, 
depending on the Rayleigh number and the range of accessible aspect ratios. A minimum of 
$Nu(\Gamma)$ is found at $\Gamma\approx 2.5$ and $\Gamma\approx 2.25$ for $Ra=10^7$ and 
$Ra=10^8$, respectively. This is exactly the point where the LSC undergoes a transition from a single-roll to a 
double-roll pattern. The trend in $Nu(\Gamma)$ curve indicates that the heat transfer becomes 
independent of the aspect ratio of the cylindrical cell for sufficiently large aspect ratios. This is 
$\Gamma\gtrsim8$ at $Ra=10^7$ and $\Gamma >8$ for $Ra\ge 10^8$. The LSC patterns reorganize 
from roll shape to pentagonal or hexagonal structures with increasing $\Gamma$ and fixed $Ra$ as 
well as with increasing $Ra$ and fixed $\Gamma$.

We provide arguments, which rationalize the non-monotonic graphs $Nu=f(\Gamma)$. Furthermore, 
we demonstrate  that the power law relation $Nu=A\times Ra^{\beta}$ gives 
rise to a coefficient $A(\Gamma)$ which decreases from 0.165 to 0.118 and an exponent 
$\beta(\Gamma)$ which increases from 0.287 to 0.305. Furthermore, they follow algebraic scaling relations 
$A(\Gamma)\sim\Gamma^{-\lambda_1}$ and $\beta(\Gamma)\sim\Gamma^{\lambda_2}$, with 
$\lambda_1=0.18$ and $\lambda_2=0.03$ for aspect ratios between 0.5 and 4 and Rayleigh 
numbers between $10^7$ and $10^9$. We believe that it is important to include this dependence, albeit 
weak, in future scaling theories. The variation of $\beta$ seems to bridge the gap between the 
well-known exponents  $\beta=$2/7 and  1/3, which have been measured in the past. Further studies 
at higher Rayleigh numbers and larger aspect ratios have to be conducted to draw a firm conclusion 
on the robustness of the observed scaling. We cannot comment on the trend with respect to Prandtl 
number, which will exist as indicated in Hartlep {\it et al.} (2005).
  
The primary POD mode  contains most of the energy, and transports about one half of 
the global heat for $\Gamma\ge1$. Their contribution to the total heat transfer varies with $\Gamma$ and 
$Ra$ as indicated in table 3. This has been demonstrated  with the help of a
Karhunen-Lo\`{e}ve analysis of samples of turbulent convection field.  We also observe 
that the LSC patterns in turbulent convection at $Ra\ge 10^7$ are still strikingly similar to those in the 
weakly nonlinear regime immediately beyond the onset of convection (Bodenschatz {\it et al.} 2000). 
The system does not seem to ``forget" these patterns. This might partly be attributed to the closed volume, 
in which the studies are conducted.  A large-scale circulation is, therefore, always present similar to 
high-Reynolds number turbulence in von K\'{a}rman swirling flows (La Porta {\it et al.} 2001) or Taylor vortex
flows (Lathrop {\it et al.} 1992). 

One possible argument against our observation of $\Gamma$--dependent Nusselt number 
could be that the Rayleigh number for the given Prandtl number $Pr=0.7$ is still too small and 
that the convective turbulence has not yet reached the so-called hard turbulence regime, 
as discussed for example by Castaing {\it et al.} (1989).  In order to weaken this argument, we 
determine the dissipation scale and relate it to the height of the cell. Since $Pr<1$, the diffusive 
scale of the temperature, the Corrsin scale $\eta_c=(\kappa^3/\langle\epsilon\rangle)
^{1/4}$, is larger than the Kolmogorov scale $\eta_K=(\nu^3/\langle\epsilon\rangle)
^{1/4}$. The scale separation ratio gives: $H/\eta_K=$133, 278 and 588 for $Ra=10^7$, $10^8$ and 
$Ra=10^9$ respectively. Here $\eta_K$ is directly evaluated from the energy 
dissipation field as discussed in section 2. Even if we take a fraction of H, the scale 
separation is of ${\cal O}(10^2)$. Furthermore, for all the Rayleigh numbers discussed here, we 
reported strongly non-Gaussian temperature statistics in Emran \& Schumacher (2008), which clearly
indicate that the convective motion is in a state of fully developed turbulence.

Further numerical simulations and experiments in the regime of large aspect ratio  and high 
Rayleigh number are necessary. One can expect that the aspect-ratio-dependence of the turbulent
heat transfer will disappear for sufficiently large $\Gamma$ and that turbulent convection approaches
an asymptotic geometric regime in which the physics becomes independent of side wall effects. 
To achieve those goals, some efforts are underway for the cylindrical case and will hopefully shed more 
light on the dependencies $A(\Gamma)$ and $\beta(\Gamma)$ in the heat transport law 
as reported  in Fig.~4. Another important aspect, in our 
view,  would be to conduct a closer study of the same issues for fixed flux boundary conditions 
(which correspond, for example, to a radiative cooling on top of an atmospheric boundary layer). 
Recently, the first step in this direction has been undertaken by Verzicco \& Sreenivasan (2007) and 
Johnston \& Doering (2008).
  
\acknowledgments
We wish to thank Roberto Verzicco for providing us his simulation code and his help
at the beginning of our studies. The authors also acknowledge the support by the Deutsche 
Forschungsgemeinschaft (DFG) under grant SCHU1410/2-1 and by the Heisenberg Program 
of the DFG under grant SCHU 1410/5-1. The largest DNS simulations have been carried 
at the J\"ulich Supercomputing Centre (Germany)  under grants HMR09 and HIL03. We thank 
F. H. Busse,  C. R. Doering, S. Grossmann, K. R. Sreenivasan,  A. Thess and K.-Q. Xia for 
helpful comments and suggestions. The work is also benefitted  from the constructive comments 
by the three anonymous referees.

\end{document}